\documentclass[10pt,twocolumn,aps,prl,superscriptaddress,reprint]{revtex4-2}

\usepackage[latin9]{inputenc}
\usepackage{setspace}

\usepackage{bm}
\usepackage{amsmath}
\usepackage{graphicx}
\usepackage[unicode=true,
 bookmarks=false,
 breaklinks=false,pdfborder={0 0 1},colorlinks=false]
 {hyperref}
\usepackage{breakurl}

\usepackage{float}
\usepackage{threeparttable} 
\usepackage{multirow}
\usepackage{lineno}
\makeatletter
\usepackage{soul,xcolor}
\usepackage{bm}% bold maths
\usepackage{url}
\usepackage{ulem}
\makeatother
\newcommand{\be}{\begin{equation}}
\newcommand{\ee}{\end{equation}}

\newcommand{\hlt}{\textcolor{black}}

\newcommand{\EPC}{EPC}

\newcommand{\TSLS}{Ta$_3$S$_4$}
\newcommand{\TLE}{TaS$_2$}
\newcommand{\TELS}{Ta$_2$S$_4$}

\newcommand{\LQ}{$\lambda_{\boldsymbol{q}\nu}$}
\newcommand{\WQ}{$\omega_{\boldsymbol{q}\nu}$}

\newcommand{\afdw}{$\alpha^{2}F(\omega)/\omega$}

\newcommand{\CHIQ}{$\chi_{\boldsymbol{q}\nu}$}
\newcommand{\CHIQO}{$\chi_{\boldsymbol{q}\nu=1}$}
\newcommand{\CHIPQ}{$\chi^{\prime}_{\boldsymbol{q}}$}
\newcommand{\CHIQOM}{$\chi_{\boldsymbol{q}=\rm{M},\nu=1}$}
\newcommand{\CHIQOMK}{$\chi_{\boldsymbol{q}=\rm{M},\nu=1}(\boldsymbol{k})$}

\newcommand{\TC}{$T_{\mathrm{c}}$}
\newcommand{\QCDW}{$\boldsymbol{q}_{\mathrm{CDW}}$}

\newcommand\redsout{\bgroup\markoverwith{\textcolor{red}{\rule[0.5ex]{2pt}{0.7pt}}}\ULon}

\setstcolor{red}

\begin{document}

\title{ Emergent charge density wave featuring quasi-one-dimensional chains in Ta-intercalated bilayer 2$H$-TaS$_{2}$ with coexisting superconductivity}

\author{Tiantian Luo}
\affiliation{Siyuan Laboratory, Guangzhou Key Laboratory of Vacuum Coating Technologies and New Energy Materials, Department of Physics, Jinan University, Guangzhou 510632, China}	

\author{Maoping Zhang}
\affiliation{Siyuan Laboratory, Guangzhou Key Laboratory of Vacuum Coating Technologies and New Energy Materials, Department of Physics, Jinan University, Guangzhou 510632, China}	

\author{Jifu Shi}
\email{shijifu2017@126.com}
\affiliation{Siyuan Laboratory, Guangzhou Key Laboratory of Vacuum Coating Technologies and New Energy Materials, Department of Physics, Jinan University, Guangzhou 510632, China}	

\author{Feipeng Zheng}
\email{fpzheng\_phy@email.jnu.edu.cn}
\affiliation{Siyuan Laboratory, Guangzhou Key Laboratory of Vacuum Coating Technologies and New Energy Materials, Department of Physics, Jinan University, Guangzhou 510632, China}	

%\date{\today}
\begin{abstract}
Recently, intercalation emerges as an  effective  way to manipulate ground-state properties and enrich quantum phase diagrams of layered transition metal dichalcogenides (TMDCs). 
In this work, we focus on fully Ta-intercalated bilayer 2H-TaS$_{2}$ with a stoichiometry of Ta$_{3}$S$_{4}$, which has recently been experimentally synthesized. 
Based on first-principles calculations, we computationally show the suppression of an intrinsic $3\times3$ charge-density wave (CDW) in the TaS$_{2}$ layer, and the emergence of a $2\times1$ CDW in  intercalated Ta layer. 
The formation of the CDW in Ta$_{3}$S$_{4}$ is  triggered by strong electron-phonon coupling (EPC) between the $d$-like orbitals of  intercalated Ta atoms via the imaginary phonon modes at M point. 
A  2$\times$1 CDW structure is identified, featuring quasi-one-dimensional Ta chains, attributable to the competition between the CDW displacements associated with  potential CDW vectors ($\boldsymbol{q}_{\text{CDW}}$s). 
Superconductivity is found to coexist with the 2$\times$1 CDW in Ta$_{3}$S$_{4}$, with an estimated superconducting transition temperature ($T_{\mathrm{c}}$) of 3.0 K, slightly higher than that of  bilayer TaS$_{2}$. 
The \TSLS~structures of non-CDW, 2$\times$1 CDW, and $2\times$2 CDW can be switched by strain.
Our work enriches the phase diagram of TaS$_{2}$, offers a candidate material for studying  the interplay between CDW and superconductivity, and highlights  intercalation as an effective way to tune the physical properties of layered materials.
\end{abstract}
\maketitle

Transition metal dichalcogenides (TMDCs) are a class of materials in which the covalent-bonding layers are held together by van der Waals interactions along their stacking direction. 
They have received increasing attention as they possess rich quantum phases  including charge-density wave (CDW), superconductivity, magnetic ordering, etc.
Furthermore, their  interlayer couplings can be easily manipulated due to weak van der Waals interaction.  
Intercalation, through which additional atoms or molecules are inserted into the interlayer space, is one of the effective methods to manipulate the interlayer coupling and has recently been shown to serve as an efficient way to tune the ground-state properties and enrich phase diagrams of TMDCs. 
For example, alkali-metal intercalations were found to suppress CDW through electron doping~\cite{Liu2021c,Lian2017}.
Organic cations intercalated bulk NbSe$_{2}$ can  host the unique properties of both monolayer and bulk NbSe$_{2}$, with the simultaneous presence of Ising superconductivity and a high superconducting transition temperature~\cite{Zhang2022}.
Fe intercalation can induce magnetic order  in two-dimensional FeSe$_{2}$ owing to the spin-density transfer between native and intercalated Fe atoms~\cite{Huan2022}.
Intercalations can also induce superconductivity in TMDCs~\cite{Wu2021,Zheng2019c} and materials composed of main-group elements such as bilayer graphene~\cite{Toyama2022,Wang2022}.

Recently, two-dimensional systems of Ta-intercalated 2$H$-TaS$_2$ (TaS$_2$ henceforth) have been experimentally synthesized, where the Ta atoms intercalated into interlayer  space of adjacent TaS$_2$ layers in Ta-rich growing environments~\cite{Zhao2020b}. 
By adjusting the chemical potential of sulfur, various kinds of Ta-intercalated bilayer TaS$_2$ (\TELS) with different stoichiometries can be  synthesized. 
In particular, 100\% Ta-intercalated \TELS~can be realized with a stoichiometry of \TSLS,~to which the interface from monolayer TaS$_{2}$ can be  observed using scanning transmission electron microscopy. 
Interestingly, first-principles calculations further show the  development of magnetic order when the Ta-intercalated concentration is less than 50\% for both bilayer and bulk Ta$_x$S$_{y}$, indicating that Ta-interaction can modify the ground states of two-dimensional TaS$_2$. 
As \TELS~intrinsically hosts the coexistence of superconductivity and a 3$\times$3 CDW~\cite{Yang2018,Navarro-Moratalla2016}, it is very interesting to ask how the Ta intercalation will modify them. 
Furthermore, TaS$_2$ systems possess  a rich phase diagram. 
In addition to the intrinsic 3$\times$3 CDW, which coexists with superconductivity and emergent magnetic order induced by low-concentration Ta intercalations, the TaS$_2$ can further transform into 2$\times$2, 4$\times$4 CDW  and even into non-CDW phase by controlling the amount of lithium doping, charge doping and the extent of hybridization with substrates~\cite{Hall2019,Liu2021c,sanders2016crystalline,Shao2019,Lian2022}. 
Thus, it is also fascinating to ask whether the Ta-intercalation could further enrich the  phase diagram of TaS$_2$ systems.

Herein, we  study the CDW and superconductivity in Ta$_{3}$S$_{4}$ based on first-principles calculations. 
We show the suppression of the 3$\times$3 CDW in the \TELS~layer and the emergence of a 2$\times$1 CDW in the intercalated Ta layer.
The  2$\times$1 CDW structure features quasi-one-dimensional Ta chains, attributable to the strong electron-phonon coupling (EPC) arising from the intercalated Ta atoms, and the competition of potential $\boldsymbol{q}_{\text{CDW}}$s  in non-CDW Ta$_{3}$S$_{4}$, where the softest phonon is present. 
Furthermore, the \TC~of the CDW \TSLS~is estimated to be 3.0 K, slightly higher than that of \TELS, owing to the combined effects of reduced electronic states available for EPC but enhanced coupling strength arising from phonon frequency dip.
The above results suggest the coexistence of superconductivity and the 2$\times$1 CDW in \TSLS.
By applying strains, the \TSLS~can switch among the states of non-CDW, $2\times1$ CDW, and $2\times2$ CDW.
%Furthermore, the Ta-intercalation, and Fermi-surface gapping induced by the CDW lead to the substantial suppression of superconductivity compared with non-intercalated TaS$_{2}$, owing to the reduced electronic density of states at Fermi energy [$N(0)$]. 

The calculations in this work were performed by Quantum Espresso~\cite{Giannozzi2009}, VASP~\cite{Kresse1996}, and EPW~\cite{Ponce2016} packages. More details can be found in Ref.~\cite{METHOD}.

\begin{figure}
	\centering
	\includegraphics[width=76 mm]{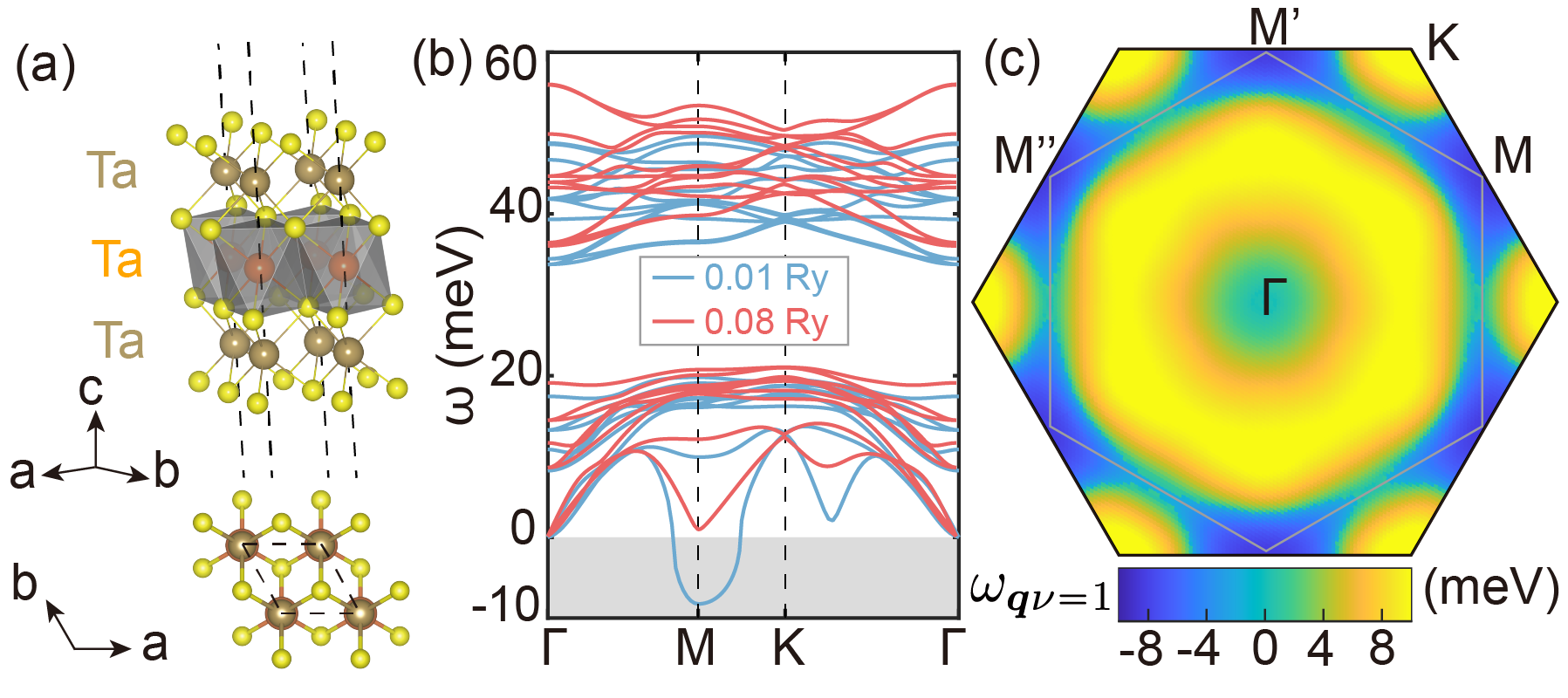} 
		\caption{Crystal structure and phonon energy of non-CDW \TSLS.
		(a) Side and top  views of the structure, where the intercalated Ta atoms are colored orange.
		(b) Phonon dispersions calculated using a regular electronic broadening ($\sigma$ = 0.01 Ry) and a large one ($\sigma$ = 0.08 Ry). 
		(c) Phonon-energy distributions of the lowest branch in BZ calculated using $\sigma = 0.01$ Ry.}
	\label{fig1}
\end{figure}

For bulk \TLE, the selected pseudopotentials and exchange-correlation functionals~\cite{METHOD} yield  in-plane and out-of-plane hexagonal lattice constants of $a$ = 3.31 \AA~ and $c$ = 12.23~\AA, respectively.
The result agrees well with experiment~in Ref.\cite{Meetsma1990}($a$ = 3.314 \AA and $c$ = 12.097~\AA), validating our computational method for the crystal structures of layered TaS$_{2}$. 
We then began by  determining the crystal structure of \TELS. 
Our calculation shows that the most energy-favorable \TELS~structure consists of $2H$ stacked two S-Ta-S monolayers with $a$ = 3.31~\AA, separated by an interlayer distance of d=2.89~\AA~(Sec.~S1~\cite{SM}). 
The monolayer consists of edge-shared TaS$_6$ triangular prisms, each formed by a Ta atom  at the prism center coordinated by 6 S atoms at the prism vertices. 
The calculated phonon dispersion ($\omega_{\boldsymbol{q}\nu}$) of \TELS~exhibits the most negative energy at 2/3$\Gamma$M (Sec.~S2, Ref.~\cite{SM}), indicating 3$\times$3 CDW instabilities, consistent with a recent experiment~\cite{Wang2022a}, which further validates our methods. 
When Ta atoms were intercalated into the interlayer space, they occupied the octahedral sites at the midpoints between the two nearest Ta atoms of adjacent monolayers~\cite{Zhao2020b}. 
When  the octahedral sites were fully occupied, 100 \% intercalated \TELS~ (\TSLS) without CDW was formed with calculated $a=3.26$ \AA~and $d=3.35$~\AA, as shown in Fig.~\ref{fig1}(a).

Interestingly, the non-CDW \TSLS~crystal is found to be dynamically unstable due to the imaginary phonon modes at three equivalent $M$ points in its Brillouin zone (BZ), which will be shown below.
There are three different $M$ points in a hexagonal BZ, i.e., $1/2\boldsymbol{a}_{1}^{*}$ (labeled as $M$), $1/2\boldsymbol{b}_{1}^{*}$ (M$^{\prime}$) and  $1/2(\boldsymbol{a}_{1}^{*} -\boldsymbol{b}_{1}^{*}$) ($M^{\prime\prime}$) with $\boldsymbol{a}_{1}^{*}$ and $\boldsymbol{b}_{1}^{*}$ being the in-plane reciprocal lattice vectors [Fig.~\ref{fig3}(b)]. 
They are equivalent for non-CDW \TSLS~due to the threefold rotational symmetry. 
As shown in Figs.~\ref*{fig1}(b) and ~\ref*{fig1}(c), the \WQ~of non-CDW \TSLS~computed using a regular electronic broadening ($\sigma = 0.01~\text{Ry}$) exhibits imaginary phonon frequencies with the most negative energies $\sim$-10 meV at the three $M$ points.
Interestingly, the displacement patterns associated with the imaginary phonon modes were found to be dominated by the in-plane vibrations of the intercalated Ta atoms (Sec.~S9, Ref.~\cite{SM}).
Furthermore, the imaginary phonon frequencies were sensitive to $\sigma$. 
When a larger $\sigma$=0.08 Ry is adopted, the imaginary phonon frequencies  are hardened to positive values with a dip at $M$ points [Fig.~\ref*{fig1}(b)], indicating that they are possibly associated with Kohn anomalies driven by EPC~\cite{Piscanec2004}. 
As $\sigma$ is positively correlated to temperature, the above result [Fig.~\ref{fig1}(b)] indicates that the phonon energies of the acoustic branch will be softened and first  become imaginary at the $M$ points as temperature decreases. 

\begin{figure}
	\centering
	\includegraphics[width=76 mm]{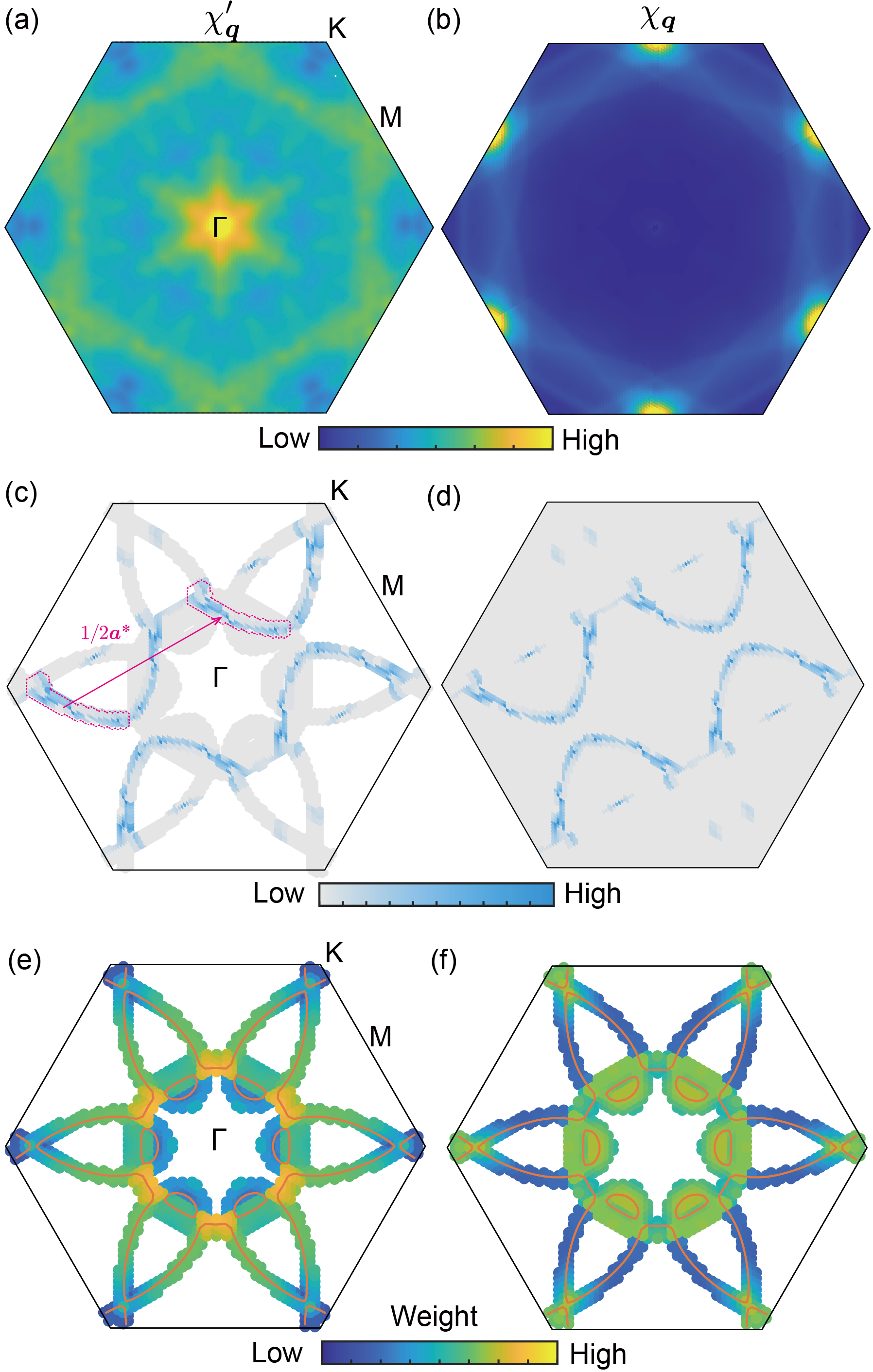} 
		\caption{Calculated quantities for non-CDW \TSLS. 
		(a) The generalized static electronic susceptibility \CHIQO. 
		(b) Bare electronic susceptibility  \CHIPQ. 
		(c-d) Distribution of \CHIQOMK, where $\mathrm{M}=1/2\boldsymbol{a}^{*}$. 
		In panel (c), those $\boldsymbol{k}$ with energies $\textgreater$0.2 eV away from Fermi energy are not shown. 
		The $\mathbf{k}$ states in the two sections with  dashed boundaries  are nested by the soft mode at $1/2\boldsymbol{a}^{*}$.
		Projected electronic states around the Fermi surface onto (e) intercalated Ta and (f) Ta in \TELS. 
		The energy window is $\pm$0.2 eV related to Fermi energy.
		The orange lines are the calculated Fermi surface.
		}
	\label{fig2}
\end{figure}

\begin{figure}
	\centering
	\includegraphics[width=76 mm]{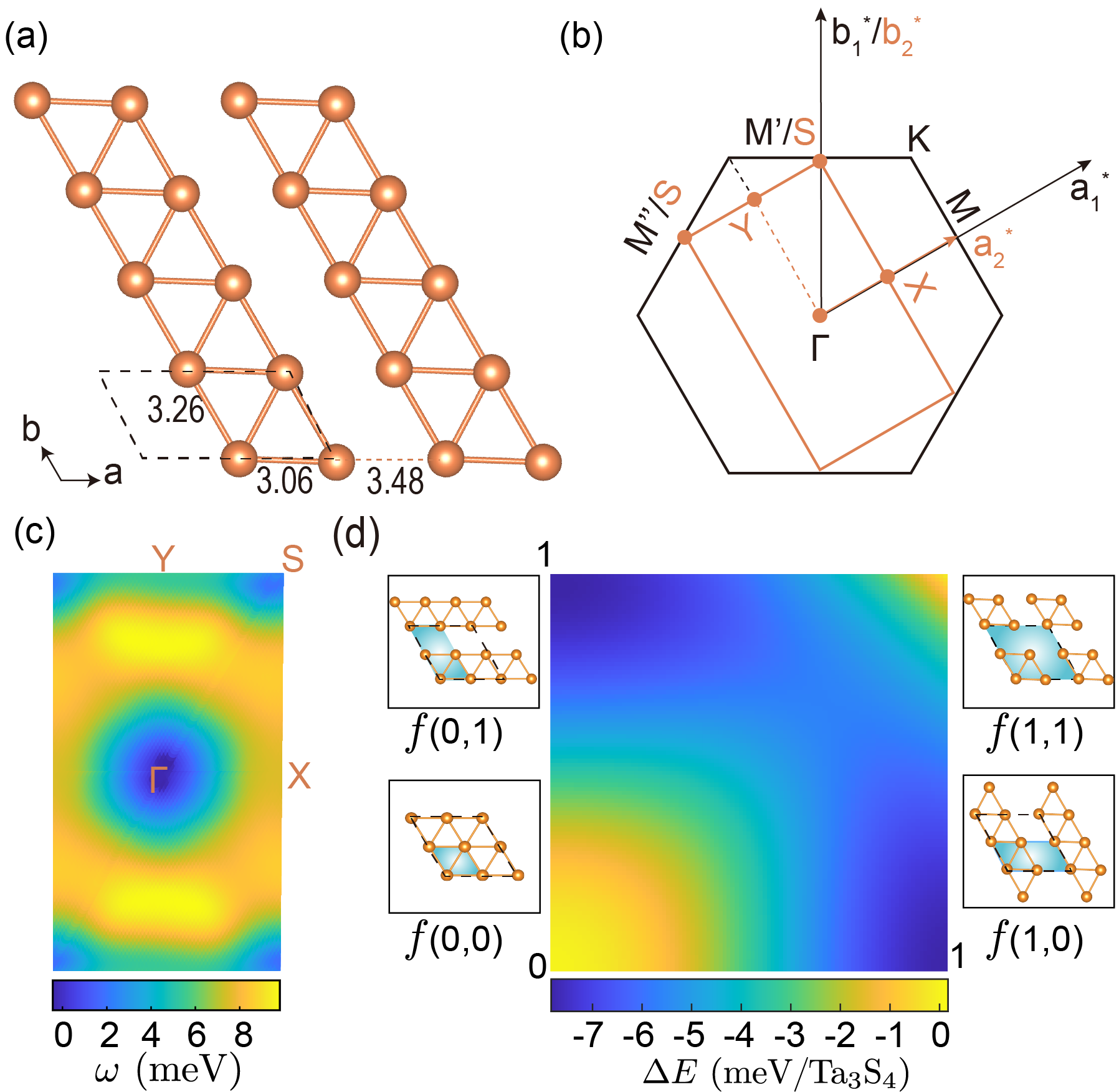} 
		\caption{ 
		(a) Top view  of the intercalated Ta-layer in the CDW \TSLS. 
		(b) The first BZs of non-CDW (black) and the CDW \TSLS~(orange), with corresponding high symmetry points. 
		(c) Distribution of $\omega_{\boldsymbol{q}\nu=1}$ for CDW \TSLS. 
		(d) Calculated energy distributions of a series of configurations in 2$\times$2 supercells, showing the energy landscape related to the development of CDW displacements associated with 1/2$\boldsymbol{a}_{1}^{*}$ and 1/2$\boldsymbol{b}_{1}^{*}$. 
		The configurations are described as $f(\alpha,\beta) = f(0,0) + g(\alpha,\beta)$, where $\alpha$ ($\beta$) $\in$ [0,1] quantifying the degree of the development of CDW displacement associated with  1/2$\boldsymbol{a}_{1}^{*}$ (1/2$\boldsymbol{b}_{1}^{*}$), and $g(\alpha,\beta) = \alpha g(1,0) + \beta  g(0,1)$ is the corresponding CDW displacement.	
		$f(0,0)$, $f(1,0)$  and $f(0,1)$ represent the non-CDW \TSLS, the optimized CDW structures related to 1/2$\boldsymbol{a}_{1}^{*}$ and 1/2$\boldsymbol{b}_{1}^{*}$, respectively, in 2$\times$2 supercells, whose intercalated layers from top view  are shown in panel (d), with colored parallelograms being their unit cells.
		}
	\label{fig3}
\end{figure}

Indeed, our calculations demonstrate that the phonon softening at $M$ points is triggered by EPC, mainly arising from intense EPC matrix elements. 
The phonon softening arising from EPC at branch $\nu$ and momentum $\boldsymbol{q}$ can be quantified by generalized static electronic susceptibility~(\CHIQ)~\cite{Giustino2017,Flicker2016,Zhu2015,Wang2022a}, which is associated with the real part of phonon self-energy due to EPC:
 \begin{equation}
    \chi_{\boldsymbol{q}\nu}=\sum_{\boldsymbol{k}, m, n} w_{\boldsymbol{k}}\left | g_{m\boldsymbol{k}, n\boldsymbol{k}+\boldsymbol{q}}^{\nu} \right |^2 \frac{f\left(\varepsilon_{m\boldsymbol{k}}\right)-f\left(\varepsilon_{n\boldsymbol{k}+\boldsymbol{q}}\right)}{\varepsilon_{n\boldsymbol{k}+\boldsymbol{q}} - \varepsilon_{m\boldsymbol{k}}},
	\label{eqCHIQ}
 \end{equation} 
 where $w_{\boldsymbol{k}}$ is the weight of the $\boldsymbol{k}$ point for BZ integration, and $f(\epsilon_{m\boldsymbol{k}})$ is the Fermi-Dirac distribution function evaluated at the electronic energy $\varepsilon_{m\boldsymbol{k}}$ associated with Kohn-Sham state ($m,\boldsymbol{k}$). 
 $g_{m\boldsymbol{k},n\boldsymbol{k}+\boldsymbol{q}}^{\nu}$ is  \EPC~matrix element quantifying the scattering amplitude between  ($m,\boldsymbol{k}$) and ($n,\boldsymbol{k}+\boldsymbol{q}$) via the phonon ($\boldsymbol{q}, \nu$).
The \CHIQ~can be further reduced to bare electronic susceptibility $\chi^{\prime}_{\boldsymbol{q}}$ in constant-matrix-element approximation:
 \begin{equation}
	\chi^{\prime}_{\boldsymbol{q}}=\sum_{\boldsymbol{k}, m, n} w_{\boldsymbol{k}} \frac{f\left(\varepsilon_{m\boldsymbol{k}}\right)-f\left(\varepsilon_{n\boldsymbol{k}+\boldsymbol{q}}\right)}{\varepsilon_{n\boldsymbol{k}+\boldsymbol{q}} - \varepsilon_{m\boldsymbol{k}}},
	\label{eqCHIPQ}
 \end{equation}
 reflecting the contributions of Fermi surface nesting~\cite{kohn1959image}, which is a pure electronic effect. 
 Therefore, the main feature of \CHIQ~is determined by combined effect of EPC matrix elements and the Fermi surface nesting. 
 To unveil the phonon softening for~\TSLS,  we computed its \CHIPQ~and \CHIQ, and the results are shown in Figs.~\ref{fig2}(a) and ~\ref{fig2}(b), respectively. 
 The \CHIPQ~features the most dominant values around $\Gamma$ owing to the intraband transitions, and the second-largest values, distributed  almost uniformly on a hexagon with six vertices at M points [Fig.~\ref*{fig2}(a)].
In comparison, the calculated \CHIQ~for the lowest phonon branch \CHIQO~features the same hexagon as \CHIPQ, but with the largest values at $M$ points [Fig.~\ref*{fig2}(b)], coinciding with the momenta where the phonon softening occurs [Fig.~\ref*{fig1}(c)]. 
It can be seen  that there already exists a moderate strength of Fermi surface nesting at $M$ points [Fig.~\ref*{fig2}(a)].
The intense EPC matrix elements  further lead to the largest value of \CHIQO~ at the $M$ points [Fig.~\ref*{fig2}(b)], which drive the phonon softening. 
The calculated $\omega_{\boldsymbol{q}\nu=1}$  also shows consistent result, where those $\boldsymbol{q}$ states on the hexagon [gray line in Fig.~\ref*{fig1}(c)] have relatively low energies. 
The above finding is similar to the cases of monolayer TaS$_2$~\cite{Wang2022a} and NbSe$_{2}$~\cite{Zheng2019a}, where the intense EPC matrix elements lead to the imaginary phonon modes at 2/3$\Gamma M$. 

To gain more insights into the appearance of the imaginary phonon modes, we then delve into  the reason for the large~\CHIQO~at $M$ points (\CHIQOM). 
The \CHIQ~can be decomposed into the contributions from each $\boldsymbol{k}$ in the BZ according to  $\chi_{\boldsymbol{q}\nu} = \sum_{\boldsymbol{k}}w_{\boldsymbol{k}}\chi_{\boldsymbol{q}\nu}(\boldsymbol{k})$. 
As the three  $M$ points are equivalent, the decomposition was only done for the $M$ at $\boldsymbol{q} = 1/2\boldsymbol{a}^{*}$ [\CHIQOMK], as displayed in Figs.~\ref*{fig2}(c) and ~\ref*{fig2}(d). 
The comparison of the above two panels shows that the dominant~\CHIQOMK~comes from the $\boldsymbol{k}$~points with the energies $\sim$0.2 eV around the Fermi level. 
To understand the orbital characters at those $\boldsymbol{k}$ points, we computed the projected electronic states around the Fermi surface as shown in Figs.~\ref*{fig2}(e), ~\ref*{fig2}(f) and Fig.~S3~\cite{SM}, along with bandstructure in Fig. 4(c) (orange lines). 
It is seen that the Fermi surface is mainly composed of the following three sections: (1) six small circles around $\Gamma$  contributed by the $d$-like orbitals of all Ta atoms; (2) a large Fermi pocket centered at $\Gamma$  with 6 leaf-shape  corners contributed mainly by the $d$-like orbitals of the intercalated Ta with a small portion of native Ta; (3) a small circle centered at each of the $K$ points mainly contributed by the $d$-like orbitals of native Ta. 
By  comparing  Fig.~\ref*{fig2}(c) with Figs.~\ref*{fig2}(e), ~\ref*{fig2}(f) and~\hlt{Fig.~S3}~\cite{SM}, one can see that the states at those $\boldsymbol{k}$ points with dominant \CHIQOMK~are mainly derived from the $d$-like orbitals of the intercalated Ta atoms around the leaf-shape section and from the hybridized $d$-like orbitals around the small circles near $\Gamma$.
Later we will show that the states involved in the CDW formation are mainly contributed by the intercalated Ta.

The $M$ points with imaginary phonon energies serve as potential \QCDW s, which are associated with reconstructed 2$\times$1 (1$\times$2) or 2$\times$2 superstructures with lower energies than the non-CDW \TSLS. 
To obtain the reconstructed structures, we built  2$\times$2  supercells, with all atoms uniformly distributed at their equilibrium positions. 
Then we introduced small random displacements on the atoms, followed by structural optimizations. 
After multiple structural optimizations, we obtained one reconstructed structure characterized by the displacements of intercalated Ta atoms with concomitant displacements of S atoms. 
The reconstructed structure features dimerizations of the Ta atoms along $\boldsymbol{a}$ direction, forming quasi-one-dimensional Ta chains along $\boldsymbol{b}$ direction in the intercalated Ta layer [Fig.~\ref*{fig3}(a)]. 
In contrast, the Ta atoms in the \TELS~layer exhibit nearly uniform distribution (Fig.~S4, Ref.~\cite{SM}), indicating the suppression of the intrinsic 3$\times$3 CDW in \TELS~(Sec. S2, Ref.~\cite{SM}). 
The above CDW displacements are consistent with the intercalated-Ta-dominated atomic vibrations at \QCDW~(Sec.~S9, Ref.~\cite{SM}), as mentioned before.
Remarkably, the reconstructed structure exhibits an equal distance of 3.26~\AA~between adjacent Ta atoms along  $\boldsymbol{b}$ direction, indicating that it possesses a smaller unit cell with a size of $2\times1$. 
This is further confirmed by direct structure optimizations  in  $2\times1$ supercells, which yield the same structure  and total energy (-7.8 meV per \TSLS~relative to non-CDW \TSLS). 
Phonon calculation was further performed to assess the dynamical stability of the reconstructed structure, and the result is shown in Fig. ~\ref*{fig5}(a). 
It can be seen that all \WQ~[Fig.~\ref*{fig5}(a)] are positive except some negligible imaginary phonon modes near $\Gamma$, indicating the dynamical stability of the structure.
The energy distribution of the lowest branch in BZ [Fig.~\ref*{fig3}(c)] further confirms the above result. 
Therefore, we have shown that the intercalants lead to the suppression of intrinsic 3$\times$3 CDW in the \TELS, and the formation of 2$\times$1 CDW in the intercalated Ta layer.

We also note that \WQ~of the CDW~\TSLS~exhibits a dip with a positive energy of 2.54 meV at $S$ points [Figs.~\ref*{fig3}(c) and \ref*{fig5}(a)], which are coincided with the $M^{\prime}$ and $M^{\prime\prime}$  in non-CDW BZ [Fig.~\ref*{fig3}(b)], where the $\omega_{\boldsymbol{q}\nu}$s are imaginary. 
This indicates the suppression of CDW instability at  $M^{\prime}$ and  $M^{\prime\prime}$, after the development of CDW associated with the $M$ point. 
This indication is confirmed by the calculations of the energy distribution of the structures in 2$\times$2 supercells generated by adding the linear combinations of CDW displacements associated with 1/2$\boldsymbol{a}_1^{*}$ and 1/2$\boldsymbol{b}_1^{*}$ to the atoms in non-CDW \TSLS~(see the caption of Fig. 3 for details). 
It can be seen in Fig.~\ref*{fig3}(d) that after the CDW displacement associated  with 1/2$\boldsymbol{a}_1^{*}$ (1/2$\boldsymbol{b}_1^{*}$) develops, along the path of (0,0)$\rightarrow$(1,0) [(0,0)$\rightarrow$(0,1)], the total energy decreases monotonically to the minimum of -7.8 meV/\TSLS~at (1,0) [(0,1)], corresponding to the CDW structures with 2$\times$1 (1$\times$2) supercells. 
However, after the full development of the 1/2$\boldsymbol{a}_1^{*}$ (1/2$\boldsymbol{b}_1^{*}$)  displacement, the 1/2$\boldsymbol{b}_1^{*}$ (1/2$\boldsymbol{a}_1^{*}$) displacement is found to be fully suppressed, as the energy increases monotonically along the path of (1,0)$\rightarrow$(1,1) [(0,1)$\rightarrow$(1,1)].
The above results indicate that in \TSLS, CDW displacements associated with different $M$ points compete with each other.
When the CDW displacement associated with one of the $M$ points develops, the displacements corresponding to the other $M$ points are suppressed. 
This is very different from layered TaS$_{2}$ without intercalation, such as monolayer TaS$_{2}$, where the cooperations of the CDW displacements associated with different \QCDW s can be seen (Fig. S5~\cite{SM}).
The above analysis gives a qualitative understanding regarding the competition among the CDW displacements related to different \QCDW s. 

\begin{figure}
	\centering
	\includegraphics[width=76 mm]{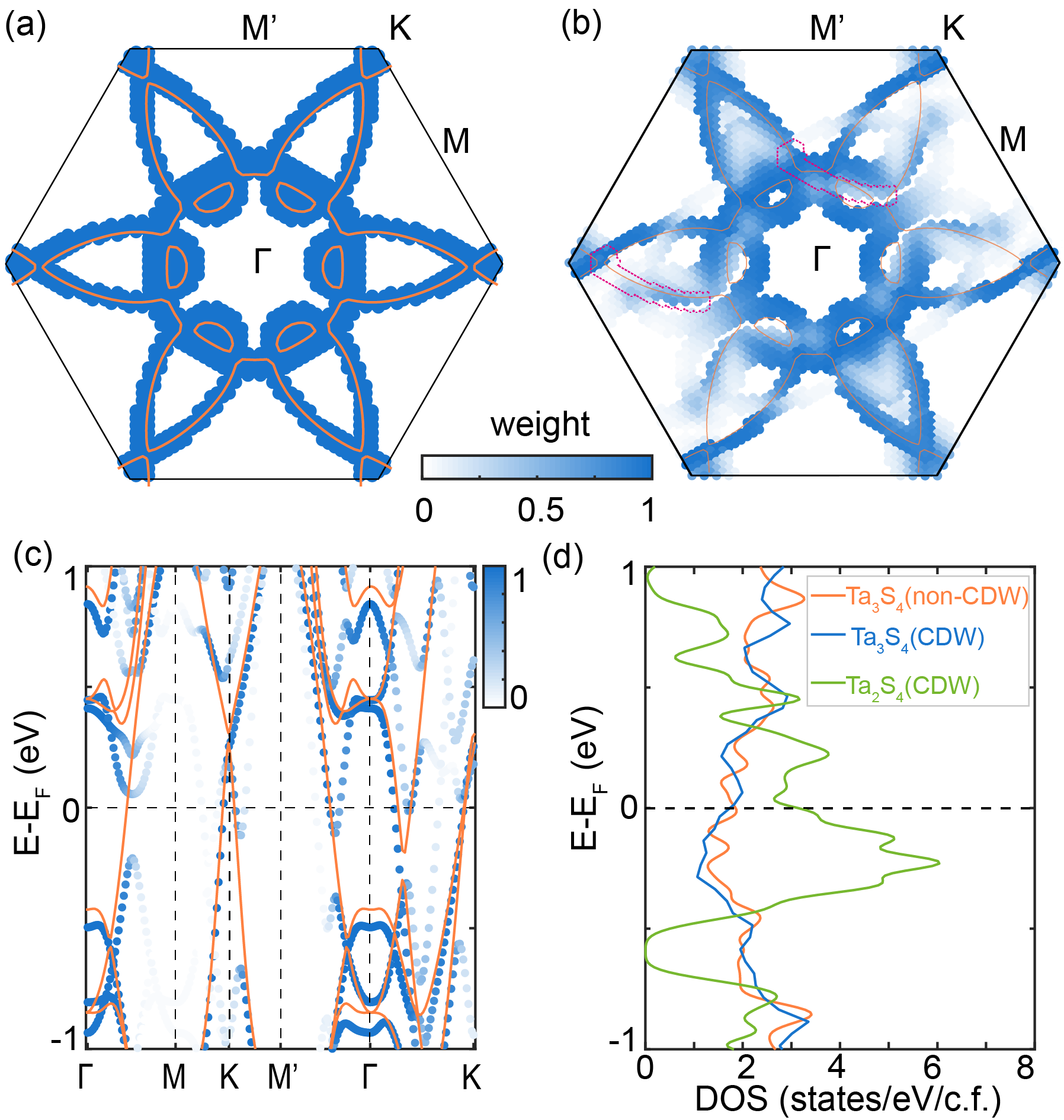} 
		\caption{Unfolded Fermi surfaces of non-CDW (a) and CDW \TSLS~(b) in 2$\times$1 supercells, where those $\boldsymbol{k}$ states with energies in $\pm$0.2  eV around Fermi energy are shown with  unfolding weights.
		The orange lines represent the Fermi surface of non-CDW \TSLS~directly calculated using its unit cell. 
		The locations of the sections with dashed boundaries are the same as that in Fig.~\ref{fig2}(c).
		(c) Unfolded bandstructure of CDW \TSLS, superimposed with the one of non-CDW \TSLS~(orange lines).
		(d) Density of states for non-CDW and CDW~\TSLS, and \TELS~ in CDW phase.
		}
	\label{fig4}
\end{figure}
After unveiling the origin of the emergent CDW in~\TSLS, an important question arises: how the electronic structure is modified by the CDW, which is essential to understand the interplay between CDW and superconductivity. 
To study the above question, we unfolded the electronic structures of CDW \TSLS~into non-CDW BZ and directly compared them with  non-CDW \TSLS~based on a BZ unfolding scheme~\cite{Allen2013,zheng2017charge}. 
The calculated unfolded Fermi surface of  non-CDW \TSLS~using a 2$\times$1 supercell is consistent with the one directly calculated using its primitive cell [Fig.~\ref*{fig4}(a)], indicating the reliability of our method. 
When the CDW forms in 2$\times$1 supercells, the Fermi surface and bandstructure are modified, as shown in Figs.~\ref*{fig4}(b) and ~\ref*{fig4}(c), respectively (see Fig.~S6~\cite{SM} for the results of three intermediate structures between non-CDW and CDW \TSLS). 
The most notable change of  the Fermi surface is the opening of gaps at the nested sections [dashed boundaries in Fig.~\ref*{fig4}(b)] with smaller unfolding weights than the non-CDW case [Fig.~\ref*{fig4}(a)].
We note that the gaps in the nested sections are not fully opened because the finite weights can be found, especially at the regions near the small circle around $\Gamma$. 
Such phenomenon is different from the cases of monolayer  NbSe$_{2}$~\cite{Zheng2019a} and TaS$_{2}$~\cite{Yang2018}, where the fully gapping at the nested sections can be found.
The above discrepancy can be due to the different orbital characters: the nested sections for the \TSLS~are contributed by all Ta atoms, and only the intercalated Ta atoms participate in the CDW displacements (see Secs.~S4 and S9~\cite{SM}, Ref.~\cite{SM}), whereas for NbSe$_{2}$, the nested sections are totally contributed by the Nb atoms, which are involved in the formation of CDW with triangle clusters.
The Fermi-surface gapping induced by the CDW reduces the density of states at Fermi level $N(0)$ from \hlt{1.85} (non-CDW) to \hlt{1.74} states/eV/Ta$_{3}$S$_{4}$ (CDW).
The gap opening can also be seen in the folded BZ, as discussed in Sec.~S13, Ref.~\cite{SM}.

Finally, as superconductivity and CDW intrinsically coexist in layered TaS$_{2}$~\cite{Navarro-Moratalla2016,Yang2018}, a remaining question is whether the superconductivity can now coexist with the  2$\times$1 CDW in ~\TSLS~after the gaps open on the Fermi surface.
Our EPC calculations show that the EPC constant for the CDW \TSLS~is $\lambda = \lambda(\infty) = 0.79$ [Fig.~\ref{fig5}(c)].
Further examination of the spectrum in Fig.~\ref{fig5}(c) shows that  the phonons contributing to the $\lambda$ comes from two energy windows at 0--20 and 34--50 meV. 
In particular, the low-energy phonons, mainly derived from Ta vibrations~[Fig.~\ref*{fig5}(b)], prevail the contribution with $\lambda(20~\mathrm{meV}) = 0.69$, corresponding to a proportion of 87.3\%. 
By further comparing Figs.~\ref*{fig5}(a)--\ref*{fig5}(c), it is seen the significant contributions mainly arise from the soft phonons with energies  $\textless$10 meV and located around $S$ point in BZ, where the  $\lambda_{\boldsymbol{q}\nu}$s are large. 
Combined with the calculated logarithmic average of phonon frequencies $\omega_{\mathrm{log}}=$ 97.3 K and a regular effective Coulomb potential $\mu^{*} = 0.15$ for TMDCs, superconducting~\TC~is estimated to be  3.0 K using Allen-Dynes-modified McMillan equation~\cite{McMillan1968,Allen1975,Giustino2017}, slightly higher than that of the \TELS~measured in an experiment (2.8 K)~\cite{Yang2018}.
This is counterintuitive at first glance, as the $N(0)$ of \TELS~is larger than that of \TSLS~[Fig.~\ref*{fig4}(d)].
Although the Ta intercalation tends to suppress $N(0)$, which reduces the number of electronic states available for EPC, their coupling strengths are enhanced due to low-energy phonons with large \LQ s~ [Figs. 5(a) and 5(c)].
The above two competitive effects lead to similar $\lambda$~\cite{TC} and \TC~between \TSLS~and \TELS.
By carefully  evaluating the spin-orbit coupling (SOC) effect in \TSLS, we find that the SOC  will not substantially influence the calculated results of non-CDW and CDW \TSLS~(Sec.~S7, Ref.~\cite{SM}). 

\begin{figure}
	\centering
	\includegraphics[width=76 mm]{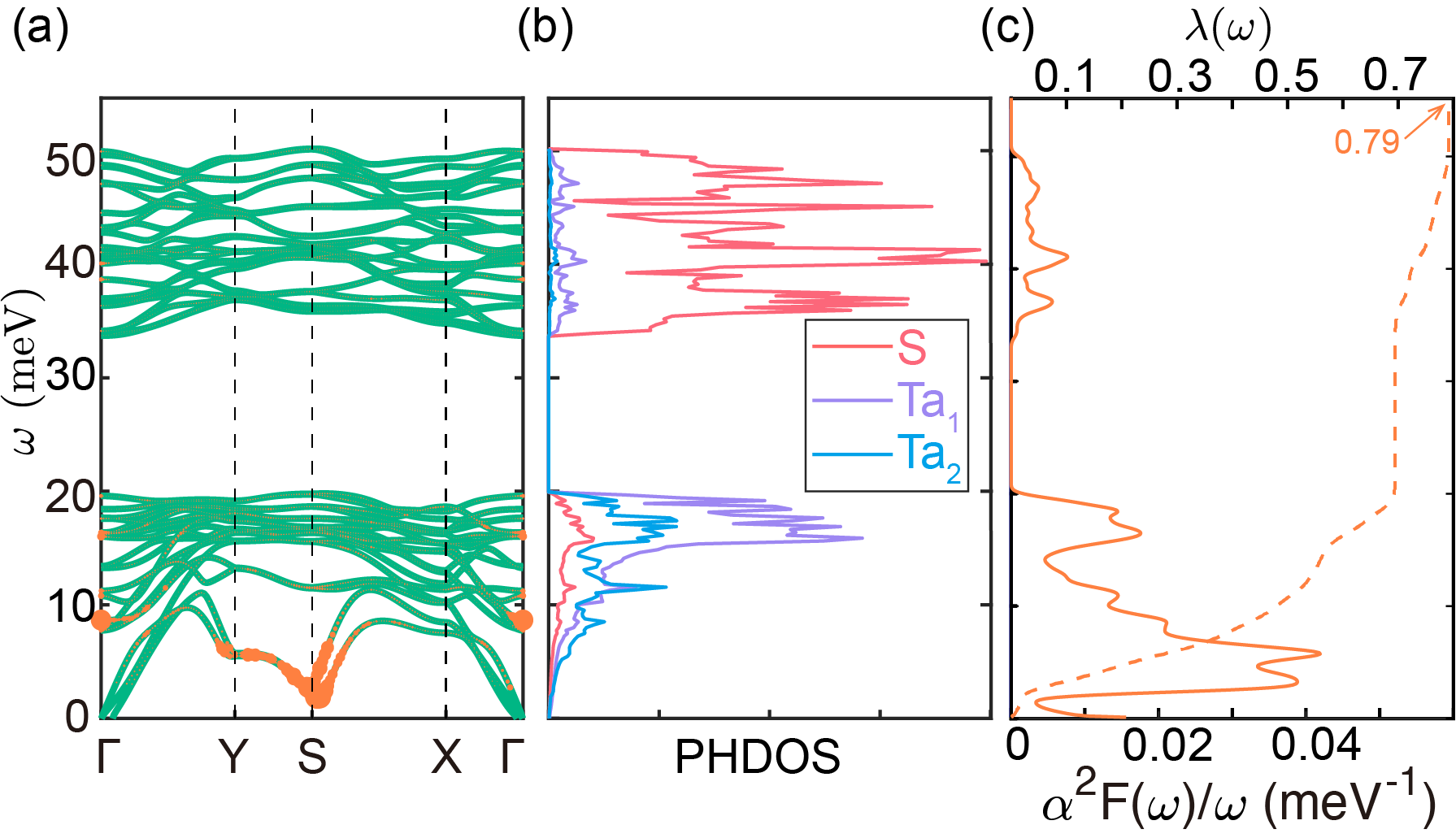} 
		\caption{
			Calculated quantities for the CDW \TSLS. 
			(a) Phonon dispersion superimposed with $\lambda_{\boldsymbol{q}\nu}$, whose values are positively correlated with the size of orange dots. 
			They were calculated using $\sigma = 0.01$ Ry. 
			(b) Projected phonon density of states onto the vibrations of S, native Ta (Ta$_{1}$), and intercalated Ta (Ta$_{2}$).
			(c) Eliashberg spectrum divided by phonon frequency [\afdw] with integrated  EPC constant [$\lambda(\omega)$].
		}
	\label{fig5}
\end{figure}

To gain more insights into the properties of \TSLS, we also study the magnetism, strain effects, and lifetime broadening of electrons and phonons, as the strain and lifetime broadening can be introduced by lattice mismatch with the substrate and fluctuations like disorders and temperatures, respectively.
The \TSLS~is calculated to be nonmagnetic (Sec.~S8, Ref.~\cite{SM}), in agreement with the previous work~\cite{Zhao2020b}. 
Furthermore, the electronic broadening will suppress the superconductivity and CDW in \TSLS. 
In particular, the CDW in \TSLS~is likely to be more robust against fluctuations than layered TaS$_{2}$~\cite{SIGMA} (Sec.~S11, Ref.~\cite{SM}).
More interestingly, the CDW and superconductivity can be effectively manipulated by strains. 
$\sim$3\% compressive strain can stabilize the CDW instability and lead to an enhanced \TC~up to $\sim$4.2 K, whereas $\sim$2.8\% tensile strain can induce a more stable 2$\times$2 CDW order~(Sec.~S10, Ref.~\cite{SM}).

In summary, we have made a computational study of crystal structures, electronic structures, EPC, and superconducting properties of \TSLS, leading to some important conclusions. 
Firstly, Ta intercalation suppresses the intrinsic 3$\times$3 CDW in \TELS, and induces 2$\times$1 CDW in~intercalated Ta layer, triggered by strong EPC of $d$-like orbitals of intercalated Ta atoms with phonons at one of the three potential \QCDW s  in non-CDW \TSLS. 
One CDW structure has been identified, featuring  quasi-one-dimensional Ta chains, owing to the competition  among the CDW displacements associated with different potential $\boldsymbol{q}_{\text{CDW}}$s. 
Furthermore, the CDW in \TSLS~leads to Fermi surface gapping in part of the leaf-shape Fermi pocket associated with the imaginary phonon mode at $M$ point, reducing $N(0)$ in CDW \TSLS.
Finally, the 2$\times$1 CDW can coexist with superconductivity with an estimated \TC~of 3.0 K, which is slightly higher than that of \TELS, attributable to the combined effects of reduced $N(0)$ but enhanced EPC coupling strength arising from the low-energy phonon modes in CDW \TSLS.
The switch among the states of non-CDW, $2\times1$, and $2\times2$ CDW for \TSLS~can be  realized by strains.

\begin{acknowledgements} 
This work is supported by National Natural Science Foundation of China 11804118, 
Guangdong Basic and Applied Basic Research Foundation (Grant No.2021A1515010041),
the Science and Technology Planning Project of Guangzhou (Grant No. 202201010222), and open project funding of  Guangzhou Key Laboratory of Vacuum Coating Technologies and New Energy Materials (KFVEKFVE20200001). 
The Calculations were performed on  high-performance computation cluster of Jinan University, and Tianhe Supercomputer System.

T. Luo and M. Zhang contribute equally to this work.
\end{acknowledgements}

\bibliographystyle{apsrev4-2}
%\bibliography{main.bib}

\begin{thebibliography}{35}%
	\makeatletter
	\providecommand \@ifxundefined [1]{%
	 \@ifx{#1\undefined}
	}%
	\providecommand \@ifnum [1]{%
	 \ifnum #1\expandafter \@firstoftwo
	 \else \expandafter \@secondoftwo
	 \fi
	}%
	\providecommand \@ifx [1]{%
	 \ifx #1\expandafter \@firstoftwo
	 \else \expandafter \@secondoftwo
	 \fi
	}%
	\providecommand \natexlab [1]{#1}%
	\providecommand \enquote  [1]{``#1''}%
	\providecommand \bibnamefont  [1]{#1}%
	\providecommand \bibfnamefont [1]{#1}%
	\providecommand \citenamefont [1]{#1}%
	\providecommand \href@noop [0]{\@secondoftwo}%
	\providecommand \href [0]{\begingroup \@sanitize@url \@href}%
	\providecommand \@href[1]{\@@startlink{#1}\@@href}%
	\providecommand \@@href[1]{\endgroup#1\@@endlink}%
	\providecommand \@sanitize@url [0]{\catcode `\\12\catcode `\$12\catcode
	  `\&12\catcode `\#12\catcode `\^12\catcode `\_12\catcode `\%12\relax}%
	\providecommand \@@startlink[1]{}%
	\providecommand \@@endlink[0]{}%
	\providecommand \url  [0]{\begingroup\@sanitize@url \@url }%
	\providecommand \@url [1]{\endgroup\@href {#1}{\urlprefix }}%
	\providecommand \urlprefix  [0]{URL }%
	\providecommand \Eprint [0]{\href }%
	\providecommand \doibase [0]{https://doi.org/}%
	\providecommand \selectlanguage [0]{\@gobble}%
	\providecommand \bibinfo  [0]{\@secondoftwo}%
	\providecommand \bibfield  [0]{\@secondoftwo}%
	\providecommand \translation [1]{[#1]}%
	\providecommand \BibitemOpen [0]{}%
	\providecommand \bibitemStop [0]{}%
	\providecommand \bibitemNoStop [0]{.\EOS\space}%
	\providecommand \EOS [0]{\spacefactor3000\relax}%
	\providecommand \BibitemShut  [1]{\csname bibitem#1\endcsname}%
	\let\auto@bib@innerbib\@empty
	%</preamble>
	\bibitem [{\citenamefont {Liu}\ \emph {et~al.}(2021)\citenamefont {Liu},
	  \citenamefont {Huangfu}, \citenamefont {Zhang}, \citenamefont {Lin},\ and\
	  \citenamefont {Schilling}}]{Liu2021c}%
	  \BibitemOpen
	  \bibfield  {author} {\bibinfo {author} {\bibfnamefont {H.}~\bibnamefont
	  {Liu}}, \bibinfo {author} {\bibfnamefont {S.}~\bibnamefont {Huangfu}},
	  \bibinfo {author} {\bibfnamefont {X.}~\bibnamefont {Zhang}}, \bibinfo
	  {author} {\bibfnamefont {H.}~\bibnamefont {Lin}},\ and\ \bibinfo {author}
	  {\bibfnamefont {A.}~\bibnamefont {Schilling}},\ }\href@noop {} {\bibfield
	  {journal} {\bibinfo  {journal} {Phys. Rev. B}\ }\textbf {\bibinfo {volume}
	  {104}},\ \bibinfo {pages} {64511} (\bibinfo {year} {2021})}\BibitemShut
	  {NoStop}%
	\bibitem [{\citenamefont {Lian}\ \emph {et~al.}(2017)\citenamefont {Lian},
	  \citenamefont {Si}, \citenamefont {Wu},\ and\ \citenamefont
	  {Duan}}]{Lian2017}%
	  \BibitemOpen
	  \bibfield  {author} {\bibinfo {author} {\bibfnamefont {C.-S.}\ \bibnamefont
	  {Lian}}, \bibinfo {author} {\bibfnamefont {C.}~\bibnamefont {Si}}, \bibinfo
	  {author} {\bibfnamefont {J.}~\bibnamefont {Wu}},\ and\ \bibinfo {author}
	  {\bibfnamefont {W.}~\bibnamefont {Duan}},\ }\href
	  {https://doi.org/10.1103/PhysRevB.96.235426} {\bibfield  {journal} {\bibinfo
	  {journal} {Phys. Rev. B}\ }\textbf {\bibinfo {volume} {96}},\ \bibinfo
	  {pages} {235426} (\bibinfo {year} {2017})}\BibitemShut {NoStop}%
	\bibitem {Zhang2022}%
	  Zhang, Haoxiong, Rousuli, Awabaikeli, Zhang, Kenan, Luo, Laipeng, Guo,   	Chenguang, Cong, Xin, Lin, Zuzhang, Bao, Changhua, Zhang, Hongyun, Xu, Shengnan, Nat. Phys., 18, 1425-1430 (2022)
	\bibitem [{\citenamefont {Huan}\ \emph {et~al.}(2022)\citenamefont {Huan},
	  \citenamefont {Luo}, \citenamefont {Han}, \citenamefont {Ge}, \citenamefont
	  {Cui}, \citenamefont {Zhu}, \citenamefont {Hu}, \citenamefont {Zheng},
	  \citenamefont {Zhao},\ and\ \citenamefont {Wang}}]{Huan2022}%
	  \BibitemOpen
	  \bibfield  {author} {\bibinfo {author} {\bibfnamefont {Y.}~\bibnamefont
	  {Huan}}, \bibinfo {author} {\bibfnamefont {T.}~\bibnamefont {Luo}}, \bibinfo
	  {author} {\bibfnamefont {X.}~\bibnamefont {Han}}, \bibinfo {author}
	  {\bibfnamefont {J.}~\bibnamefont {Ge}}, \bibinfo {author} {\bibfnamefont
	  {F.}~\bibnamefont {Cui}}, \bibinfo {author} {\bibfnamefont {L.}~\bibnamefont
	  {Zhu}}, \bibinfo {author} {\bibfnamefont {J.}~\bibnamefont {Hu}}, \bibinfo
	  {author} {\bibfnamefont {F.}~\bibnamefont {Zheng}}, \bibinfo {author}
	  {\bibfnamefont {X.}~\bibnamefont {Zhao}}, \bibinfo {author}
	  {\bibfnamefont {L.}~\bibnamefont {Wang}}, \bibinfo {author}
	  {\bibfnamefont {J.}~\bibnamefont {Wang}}, \ and\ \bibinfo {author}
	  {\bibfnamefont {Y.}~\bibnamefont {Zhang}},\ }\href
	  {https://doi.org/doi.org/10.1002/adma.202207276} {\bibfield  {journal}
	  {\bibinfo  {journal} {Adv. Mater.}\ \textbf {\bibinfo {volume} {35}},\ \bibinfo {pages} {2207276}} (\bibinfo
	  {year} {2023})}\BibitemShut {NoStop}%
	\bibitem [{\citenamefont {Wu}\ \emph {et~al.}(2021)\citenamefont {Wu},
	  \citenamefont {Lin}, \citenamefont {Xiong}, \citenamefont {Li}, \citenamefont
	  {Luo}, \citenamefont {Chen},\ and\ \citenamefont {Zheng}}]{Wu2021}%
	  \BibitemOpen
	  \bibfield  {author} {\bibinfo {author} {\bibfnamefont {D.}~\bibnamefont
	  {Wu}}, \bibinfo {author} {\bibfnamefont {Y.}~\bibnamefont {Lin}}, \bibinfo
	  {author} {\bibfnamefont {L.}~\bibnamefont {Xiong}}, \bibinfo {author}
	  {\bibfnamefont {J.}~\bibnamefont {Li}}, \bibinfo {author} {\bibfnamefont
	  {T.}~\bibnamefont {Luo}}, \bibinfo {author} {\bibfnamefont {D.}~\bibnamefont
	  {Chen}},\ and\ \bibinfo {author} {\bibfnamefont {F.}~\bibnamefont {Zheng}},\
	  }\href {https://doi.org/10.1103/PhysRevB.103.224502} {\bibfield  {journal}
	  {\bibinfo  {journal} {Phys. Rev. B}\ }\textbf {\bibinfo {volume} {103}},\
	  \bibinfo {pages} {224502} (\bibinfo {year} {2021})}\BibitemShut {NoStop}%
	\bibitem [{\citenamefont {Zheng}\ \emph {et~al.}(2020)\citenamefont {Zheng},
	  \citenamefont {Li}, \citenamefont {Tan}, \citenamefont {Lin}, \citenamefont
	  {Xiong}, \citenamefont {Chen},\ and\ \citenamefont {Feng}}]{Zheng2019c}%
	  \BibitemOpen
	  \bibfield  {author} {\bibinfo {author} {\bibfnamefont {F.}~\bibnamefont
	  {Zheng}}, \bibinfo {author} {\bibfnamefont {X.-B.}\ \bibnamefont {Li}},
	  \bibinfo {author} {\bibfnamefont {P.}~\bibnamefont {Tan}}, \bibinfo {author}
	  {\bibfnamefont {Y.}~\bibnamefont {Lin}}, \bibinfo {author} {\bibfnamefont
	  {L.}~\bibnamefont {Xiong}}, \bibinfo {author} {\bibfnamefont
	  {X.}~\bibnamefont {Chen}},\ and\ \bibinfo {author} {\bibfnamefont
	  {J.}~\bibnamefont {Feng}},\ }\href
	  {https://doi.org/10.1103/PhysRevB.101.100505} {\bibfield  {journal} {\bibinfo
	   {journal} {Phys. Rev. B}\ }\textbf {\bibinfo {volume} {101}},\ \bibinfo
	  {pages} {100505(R)} (\bibinfo {year} {2020})}\BibitemShut {NoStop}%
	\bibitem [{\citenamefont {Toyama}\ \emph {et~al.}(2022)\citenamefont {Toyama},
	  \citenamefont {Akiyama}, \citenamefont {Ichinokura}, \citenamefont
	  {Hashizume}, \citenamefont {Iimori}, \citenamefont {Endo}, \citenamefont
	  {Hobara}, \citenamefont {Matsui}, \citenamefont {Horii}, \citenamefont
	  {Sato}, \citenamefont {Hirahara}, \citenamefont {Komori},\ and\ \citenamefont
	  {Hasegawa}}]{Toyama2022}%
	  \BibitemOpen
	  \bibfield  {author} {\bibinfo {author} {\bibfnamefont {H.}~\bibnamefont
	  {Toyama}}, \bibinfo {author} {\bibfnamefont {R.}~\bibnamefont {Akiyama}},
	  \bibinfo {author} {\bibfnamefont {S.}~\bibnamefont {Ichinokura}}, \bibinfo
	  {author} {\bibfnamefont {M.}~\bibnamefont {Hashizume}}, \bibinfo {author}
	  {\bibfnamefont {T.}~\bibnamefont {Iimori}}, \bibinfo {author} {\bibfnamefont
	  {Y.}~\bibnamefont {Endo}}, \bibinfo {author} {\bibfnamefont {R.}~\bibnamefont
	  {Hobara}}, \bibinfo {author} {\bibfnamefont {T.}~\bibnamefont {Matsui}},
	  \bibinfo {author} {\bibfnamefont {K.}~\bibnamefont {Horii}}, \bibinfo
	  {author} {\bibfnamefont {S.}~\bibnamefont {Sato}}, \bibinfo {author}
	  {\bibfnamefont {T.}~\bibnamefont {Hirahara}}, \bibinfo {author}
	  {\bibfnamefont {F.}~\bibnamefont {Komori}},\ and\ \bibinfo {author}
	  {\bibfnamefont {S.}~\bibnamefont {Hasegawa}},\ }\href
	  {https://doi.org/10.1021/acsnano.1c11161} {\bibfield  {journal} {\bibinfo
	  {journal} {ACS Nano}\ }\textbf {\bibinfo {volume} {16}},\ \bibinfo {pages}
	  {3582} (\bibinfo {year} {2022})}\BibitemShut {NoStop}%
	\bibitem [{\citenamefont {Wang}\ \emph
	  {et~al.}(2022{\natexlab{a}})\citenamefont {Wang}, \citenamefont {Liu},
	  \citenamefont {Wu}, \citenamefont {Qu}, \citenamefont {Zhang}, \citenamefont
	  {Wang}, \citenamefont {Guan}, \citenamefont {Wang}, \citenamefont {Zheng},
	  \citenamefont {Li}, \citenamefont {Liu},\ and\ \citenamefont
	  {Jia}}]{Wang2022}%
	  \BibitemOpen
	  \bibfield  {author} {\bibinfo {author} {\bibfnamefont {X.}~\bibnamefont
	  {Wang}}, \bibinfo {author} {\bibfnamefont {N.}~\bibnamefont {Liu}}, \bibinfo
	  {author} {\bibfnamefont {Y.}~\bibnamefont {Wu}}, \bibinfo {author}
	  {\bibfnamefont {Y.}~\bibnamefont {Qu}}, \bibinfo {author} {\bibfnamefont
	  {W.}~\bibnamefont {Zhang}}, \bibinfo {author} {\bibfnamefont
	  {J.}~\bibnamefont {Wang}}, \bibinfo {author} {\bibfnamefont {D.}~\bibnamefont
	  {Guan}}, \bibinfo {author} {\bibfnamefont {S.}~\bibnamefont {Wang}}, \bibinfo
	  {author} {\bibfnamefont {H.}~\bibnamefont {Zheng}}, \bibinfo {author}
	  {\bibfnamefont {Y.}~\bibnamefont {Li}}, \bibinfo {author} {\bibfnamefont
	  {C.}~\bibnamefont {Liu}},\ and\ \bibinfo {author} {\bibfnamefont
	  {J.}~\bibnamefont {Jia}},\ }\href
	  {https://doi.org/10.1021/acs.nanolett.2c02804} {\bibfield  {journal}
	  {\bibinfo  {journal} {Nano Lett.}\ }\textbf {\bibinfo {volume} {22}},\
	  \bibinfo {pages} {7651} (\bibinfo {year} {2022}{\natexlab{a}})}\BibitemShut
	  {NoStop}%
	\bibitem [{\citenamefont {Zhao}\ \emph {et~al.}(2020)\citenamefont {Zhao},
	  \citenamefont {Song}, \citenamefont {Wang}, \citenamefont {Riis-Jensen},
	  \citenamefont {Fu}, \citenamefont {Deng}, \citenamefont {Wan}, \citenamefont
	  {Kang}, \citenamefont {Ning}, \citenamefont {Dan}, \citenamefont
	  {Venkatesan}, \citenamefont {Liu}, \citenamefont {Zhou}, \citenamefont
	  {Thygesen}, \citenamefont {Luo}, \citenamefont {Pennycook},\ and\
	  \citenamefont {Loh}}]{Zhao2020b}%
	  \BibitemOpen
	  \bibfield  {author} {\bibinfo {author} {\bibfnamefont {X.}~\bibnamefont
	  {Zhao}}, \bibinfo {author} {\bibfnamefont {P.}~\bibnamefont {Song}}, \bibinfo
	  {author} {\bibfnamefont {C.}~\bibnamefont {Wang}}, \bibinfo {author}
	  {\bibfnamefont {A.~C.}\ \bibnamefont {Riis-Jensen}}, \bibinfo {author}
	  {\bibfnamefont {W.}~\bibnamefont {Fu}}, \bibinfo {author} {\bibfnamefont
	  {Y.}~\bibnamefont {Deng}}, \bibinfo {author} {\bibfnamefont {D.}~\bibnamefont
	  {Wan}}, \bibinfo {author} {\bibfnamefont {L.}~\bibnamefont {Kang}}, \bibinfo
	  {author} {\bibfnamefont {S.}~\bibnamefont {Ning}}, \bibinfo {author}
	  {\bibfnamefont {J.}~\bibnamefont {Dan}}, \bibinfo {author} {\bibfnamefont
	  {T.}~\bibnamefont {Venkatesan}}, \bibinfo {author} {\bibfnamefont
	  {Z.}~\bibnamefont {Liu}}, \bibinfo {author} {\bibfnamefont {W.}~\bibnamefont
	  {Zhou}}, \bibinfo {author} {\bibfnamefont {K.~S.}\ \bibnamefont {Thygesen}},
	  \bibinfo {author} {\bibfnamefont {X.}~\bibnamefont {Luo}}, \bibinfo {author}
	  {\bibfnamefont {S.~J.}\ \bibnamefont {Pennycook}},\ and\ \bibinfo {author}
	  {\bibfnamefont {K.~P.}\ \bibnamefont {Loh}},\ }\href
	  {https://doi.org/10.1038/s41586-020-2241-9} {\bibfield  {journal} {\bibinfo
	  {journal} {Nature}\ }\textbf {\bibinfo {volume} {581}},\ \bibinfo {pages}
	  {171} (\bibinfo {year} {2020})}\BibitemShut {NoStop}%
	\bibitem [{\citenamefont {Yang}\ \emph {et~al.}(2018)\citenamefont {Yang},
	  \citenamefont {Fang}, \citenamefont {Fatemi}, \citenamefont {Ruhman},
	  \citenamefont {Navarro-Moratalla}, \citenamefont {Watanabe}, \citenamefont
	  {Taniguchi}, \citenamefont {Kaxiras},\ and\ \citenamefont
	  {Jarillo-Herrero}}]{Yang2018}%
	  \BibitemOpen
	  \bibfield  {author} {\bibinfo {author} {\bibfnamefont {Y.}~\bibnamefont
	  {Yang}}, \bibinfo {author} {\bibfnamefont {S.}~\bibnamefont {Fang}}, \bibinfo
	  {author} {\bibfnamefont {V.}~\bibnamefont {Fatemi}}, \bibinfo {author}
	  {\bibfnamefont {J.}~\bibnamefont {Ruhman}}, \bibinfo {author} {\bibfnamefont
	  {E.}~\bibnamefont {Navarro-Moratalla}}, \bibinfo {author} {\bibfnamefont
	  {K.}~\bibnamefont {Watanabe}}, \bibinfo {author} {\bibfnamefont
	  {T.}~\bibnamefont {Taniguchi}}, \bibinfo {author} {\bibfnamefont
	  {E.}~\bibnamefont {Kaxiras}},\ and\ \bibinfo {author} {\bibfnamefont
	  {P.}~\bibnamefont {Jarillo-Herrero}},\ }\href
	  {https://doi.org/10.1103/PhysRevB.98.035203} {\bibfield  {journal} {\bibinfo
	  {journal} {Phys. Rev. B}\ }\textbf {\bibinfo {volume} {98}},\ \bibinfo
	  {pages} {35203} (\bibinfo {year} {2018})}\BibitemShut {NoStop}%
	\bibitem [{\citenamefont {Navarro-Moratalla}\ \emph {et~al.}(2016)\citenamefont
	  {Navarro-Moratalla}, \citenamefont {Island}, \citenamefont
	  {Man{\~{a}}s-Valero}, \citenamefont {Pinilla-Cienfuegos}, \citenamefont
	  {Castellanos-Gomez}, \citenamefont {Quereda}, \citenamefont
	  {Rubio-Bollinger}, \citenamefont {Chirolli}, \citenamefont
	  {Silva-Guill{\'{e}}n}, \citenamefont {Agra{\"{i}}t}, \citenamefont {Steele},
	  \citenamefont {Guinea}, \citenamefont {{Van Der Zant}},\ and\ \citenamefont
	  {Coronado}}]{Navarro-Moratalla2016}%
	  \BibitemOpen
	  \bibfield  {author} {\bibinfo {author} {\bibfnamefont {E.}~\bibnamefont
	  {Navarro-Moratalla}}, \bibinfo {author} {\bibfnamefont {J.~O.}\ \bibnamefont
	  {Island}}, \bibinfo {author} {\bibfnamefont {S.}~\bibnamefont
	  {Man{\~{a}}s-Valero}}, \bibinfo {author} {\bibfnamefont {E.}~\bibnamefont
	  {Pinilla-Cienfuegos}}, \bibinfo {author} {\bibfnamefont {A.}~\bibnamefont
	  {Castellanos-Gomez}}, \bibinfo {author} {\bibfnamefont {J.}~\bibnamefont
	  {Quereda}}, \bibinfo {author} {\bibfnamefont {G.}~\bibnamefont
	  {Rubio-Bollinger}}, \bibinfo {author} {\bibfnamefont {L.}~\bibnamefont
	  {Chirolli}}, \bibinfo {author} {\bibfnamefont {J.~A.}\ \bibnamefont
	  {Silva-Guill{\'{e}}n}}, \bibinfo {author} {\bibfnamefont {N.}~\bibnamefont
	  {Agra{\"{i}}t}}, \bibinfo {author} {\bibfnamefont {G.~A.}\ \bibnamefont
	  {Steele}}, \bibinfo {author} {\bibfnamefont {F.}~\bibnamefont {Guinea}},
	  \bibinfo {author} {\bibfnamefont {H.~S.}\ \bibnamefont {{Van Der Zant}}},\
	  and\ \bibinfo {author} {\bibfnamefont {E.}~\bibnamefont {Coronado}},\ }\href
	  {https://doi.org/10.1038/ncomms11043} {\bibfield  {journal} {\bibinfo
	  {journal} {Nat. Commun.}\ }\textbf {\bibinfo {volume} {7}},\ \bibinfo {pages}
	  {11043} (\bibinfo {year} {2016})}\BibitemShut {NoStop}%
	\bibitem [{\citenamefont {Hall}\ \emph {et~al.}(2019)\citenamefont {Hall},
	  \citenamefont {Ehlen}, \citenamefont {Berges}, \citenamefont {{Van Loon}},
	  \citenamefont {{Van Efferen}}, \citenamefont {Murray}, \citenamefont
	  {R{\"{o}}sner}, \citenamefont {Li}, \citenamefont {Senkovskiy}, \citenamefont
	  {Hell}, \citenamefont {Rolf}, \citenamefont {Heider}, \citenamefont
	  {Asensio}, \citenamefont {Avila}, \citenamefont {Plucinski}, \citenamefont
	  {Wehling}, \citenamefont {Gr{\"{u}}neis},\ and\ \citenamefont
	  {Michely}}]{Hall2019}%
	  \BibitemOpen
	  \bibfield  {author} {\bibinfo {author} {\bibfnamefont {J.}~\bibnamefont
	  {Hall}}, \bibinfo {author} {\bibfnamefont {N.}~\bibnamefont {Ehlen}},
	  \bibinfo {author} {\bibfnamefont {J.}~\bibnamefont {Berges}}, \bibinfo
	  {author} {\bibfnamefont {E.}~\bibnamefont {{Van Loon}}}, \bibinfo {author}
	  {\bibfnamefont {C.}~\bibnamefont {{Van Efferen}}}, \bibinfo {author}
	  {\bibfnamefont {C.}~\bibnamefont {Murray}}, \bibinfo {author} {\bibfnamefont
	  {M.}~\bibnamefont {R{\"{o}}sner}}, \bibinfo {author} {\bibfnamefont
	  {J.}~\bibnamefont {Li}}, \bibinfo {author} {\bibfnamefont {B.~V.}\
	  \bibnamefont {Senkovskiy}}, \bibinfo {author} {\bibfnamefont
	  {M.}~\bibnamefont {Hell}}, \bibinfo {author} {\bibfnamefont {M.}~\bibnamefont
	  {Rolf}}, \bibinfo {author} {\bibfnamefont {T.}~\bibnamefont {Heider}},
	  \bibinfo {author} {\bibfnamefont {M.~C.}\ \bibnamefont {Asensio}}, \bibinfo
	  {author} {\bibfnamefont {J.}~\bibnamefont {Avila}}, \bibinfo {author}
	  {\bibfnamefont {L.}~\bibnamefont {Plucinski}}, \bibinfo {author}
	  {\bibfnamefont {T.}~\bibnamefont {Wehling}}, \bibinfo {author} {\bibfnamefont
	  {A.}~\bibnamefont {Gr{\"{u}}neis}},\ and\ \bibinfo {author} {\bibfnamefont
	  {T.}~\bibnamefont {Michely}},\ }\href
	  {https://doi.org/10.1021/acsnano.9b03419} {\bibfield  {journal} {\bibinfo
	  {journal} {ACS Nano}\ }\textbf {\bibinfo {volume} {13}},\ \bibinfo {pages}
	  {10210} (\bibinfo {year} {2019})}\BibitemShut {NoStop}%
	\bibitem [{\citenamefont {Sanders}\ \emph {et~al.}(2016)\citenamefont
	  {Sanders}, \citenamefont {Dendzik}, \citenamefont {Ngankeu}, \citenamefont
	  {Eich}, \citenamefont {Bruix}, \citenamefont {Bianchi}, \citenamefont {Miwa},
	  \citenamefont {Hammer}, \citenamefont {Khajetoorians},\ and\ \citenamefont
	  {Hofmann}}]{sanders2016crystalline}%
	  \BibitemOpen
	  \bibfield  {author} {\bibinfo {author} {\bibfnamefont {C.~E.}\ \bibnamefont
	  {Sanders}}, \bibinfo {author} {\bibfnamefont {M.}~\bibnamefont {Dendzik}},
	  \bibinfo {author} {\bibfnamefont {A.~S.}\ \bibnamefont {Ngankeu}}, \bibinfo
	  {author} {\bibfnamefont {A.}~\bibnamefont {Eich}}, \bibinfo {author}
	  {\bibfnamefont {A.}~\bibnamefont {Bruix}}, \bibinfo {author} {\bibfnamefont
	  {M.}~\bibnamefont {Bianchi}}, \bibinfo {author} {\bibfnamefont {J.~A.}\
	  \bibnamefont {Miwa}}, \bibinfo {author} {\bibfnamefont {B.}~\bibnamefont
	  {Hammer}}, \bibinfo {author} {\bibfnamefont {A.~A.}\ \bibnamefont
	  {Khajetoorians}},\ and\ \bibinfo {author} {\bibfnamefont {P.}~\bibnamefont
	  {Hofmann}},\ }\href@noop {} {\bibfield  {journal} {\bibinfo  {journal} {Phys.
	  Rev. B}\ }\textbf {\bibinfo {volume} {94}},\ \bibinfo {pages} {81404}
	  (\bibinfo {year} {2016})}\BibitemShut {NoStop}%
	\bibitem [{\citenamefont {Shao}\ \emph {et~al.}(2019)\citenamefont {Shao},
	  \citenamefont {Eich}, \citenamefont {Sanders}, \citenamefont {Ngankeu},
	  \citenamefont {Bianchi}, \citenamefont {Hofmann}, \citenamefont
	  {Khajetoorians},\ and\ \citenamefont {Wehling}}]{Shao2019}%
	  \BibitemOpen
	  \bibfield  {author} {\bibinfo {author} {\bibfnamefont {B.}~\bibnamefont
	  {Shao}}, \bibinfo {author} {\bibfnamefont {A.}~\bibnamefont {Eich}}, \bibinfo
	  {author} {\bibfnamefont {C.}~\bibnamefont {Sanders}}, \bibinfo {author}
	  {\bibfnamefont {A.~S.}\ \bibnamefont {Ngankeu}}, \bibinfo {author}
	  {\bibfnamefont {M.}~\bibnamefont {Bianchi}}, \bibinfo {author} {\bibfnamefont
	  {P.}~\bibnamefont {Hofmann}}, \bibinfo {author} {\bibfnamefont {A.~A.}\
	  \bibnamefont {Khajetoorians}},\ and\ \bibinfo {author} {\bibfnamefont
	  {T.~O.}\ \bibnamefont {Wehling}},\ }\href
	  {https://doi.org/10.1038/s41467-018-08088-8} {\bibfield  {journal} {\bibinfo
	  {journal} {Nat. Commun.}\ }\textbf {\bibinfo {volume} {10}},\ \bibinfo
	  {pages} {180} (\bibinfo {year} {2019})}\BibitemShut {NoStop}%
	\bibitem [{\citenamefont {Lian}\ \emph {et~al.}(2022)\citenamefont {Lian},
	  \citenamefont {Heil}, \citenamefont {Liu}, \citenamefont {Si}, \citenamefont
	  {Giustino},\ and\ \citenamefont {Duan}}]{Lian2022}%
	  \BibitemOpen
	  \bibfield  {author} {\bibinfo {author} {\bibfnamefont {C.~S.}\ \bibnamefont
	  {Lian}}, \bibinfo {author} {\bibfnamefont {C.}~\bibnamefont {Heil}}, \bibinfo
	  {author} {\bibfnamefont {X.}~\bibnamefont {Liu}}, \bibinfo {author}
	  {\bibfnamefont {C.}~\bibnamefont {Si}}, \bibinfo {author} {\bibfnamefont
	  {F.}~\bibnamefont {Giustino}},\ and\ \bibinfo {author} {\bibfnamefont
	  {W.}~\bibnamefont {Duan}},\ }\href
	  {https://doi.org/10.1103/PhysRevB.105.L180505} {\bibfield  {journal}
	  {\bibinfo  {journal} {Phys. Rev. B}\ }\textbf {\bibinfo {volume} {105}},\
	  \bibinfo {pages} {L180505} (\bibinfo {year} {2022})}\BibitemShut {NoStop}%
	\bibitem [{\citenamefont {Giannozzi}\ \emph {et~al.}(2009)\citenamefont
	  {Giannozzi}, \citenamefont {Baroni}, \citenamefont {Bonini}, \citenamefont
	  {Calandra}, \citenamefont {Car}, \citenamefont {Cavazzoni}, \citenamefont
	  {Ceresoli}, \citenamefont {Chiarotti}, \citenamefont {Cococcioni},
	  \citenamefont {Dabo}, \citenamefont {{Dal Corso}}, \citenamefont
	  {de~Gironcoli}, \citenamefont {Fabris}, \citenamefont {Fratesi},
	  \citenamefont {Gebauer}, \citenamefont {Gerstmann}, \citenamefont
	  {Gougoussis}, \citenamefont {Kokalj}, \citenamefont {Lazzeri}, \citenamefont
	  {Martin-Samos}, \citenamefont {Marzari}, \citenamefont {Mauri}, \citenamefont
	  {Mazzarello}, \citenamefont {Paolini}, \citenamefont {Pasquarello},
	  \citenamefont {Paulatto}, \citenamefont {Sbraccia}, \citenamefont {Scandolo},
	  \citenamefont {Sclauzero}, \citenamefont {Seitsonen}, \citenamefont
	  {Smogunov}, \citenamefont {Umari},\ and\ \citenamefont
	  {Wentzcovitch}}]{Giannozzi2009}%
	  \BibitemOpen
	  \bibfield  {author} {\bibinfo {author} {\bibfnamefont {P.}~\bibnamefont
	  {Giannozzi}}, \bibinfo {author} {\bibfnamefont {S.}~\bibnamefont {Baroni}},
	  \bibinfo {author} {\bibfnamefont {N.}~\bibnamefont {Bonini}}, \bibinfo
	  {author} {\bibfnamefont {M.}~\bibnamefont {Calandra}}, \bibinfo {author}
	  {\bibfnamefont {R.}~\bibnamefont {Car}}, \bibinfo {author} {\bibfnamefont
	  {C.}~\bibnamefont {Cavazzoni}}, \bibinfo {author} {\bibfnamefont
	  {D.}~\bibnamefont {Ceresoli}}, \bibinfo {author} {\bibfnamefont {G.~L.}\
	  \bibnamefont {Chiarotti}}, \bibinfo {author} {\bibfnamefont {M.}~\bibnamefont
	  {Cococcioni}}, \bibinfo {author} {\bibfnamefont {I.}~\bibnamefont {Dabo}},
	  \bibinfo {author} {\bibfnamefont {A.}~\bibnamefont {{Dal Corso}}}, \bibinfo
	  {author} {\bibfnamefont {S.}~\bibnamefont {de~Gironcoli}}, \bibinfo {author}
	  {\bibfnamefont {S.}~\bibnamefont {Fabris}}, \bibinfo {author} {\bibfnamefont
	  {G.}~\bibnamefont {Fratesi}}, \bibinfo {author} {\bibfnamefont
	  {R.}~\bibnamefont {Gebauer}}, \bibinfo {author} {\bibfnamefont
	  {U.}~\bibnamefont {Gerstmann}}, \bibinfo {author} {\bibfnamefont
	  {C.}~\bibnamefont {Gougoussis}}, \bibinfo {author} {\bibfnamefont
	  {A.}~\bibnamefont {Kokalj}}, \bibinfo {author} {\bibfnamefont
	  {M.}~\bibnamefont {Lazzeri}}, \bibinfo {author} {\bibfnamefont
	  {L.}~\bibnamefont {Martin-Samos}}, \bibinfo {author} {\bibfnamefont
	  {N.}~\bibnamefont {Marzari}}, \bibinfo {author} {\bibfnamefont
	  {F.}~\bibnamefont {Mauri}}, \bibinfo {author} {\bibfnamefont
	  {R.}~\bibnamefont {Mazzarello}}, \bibinfo {author} {\bibfnamefont
	  {S.}~\bibnamefont {Paolini}}, \bibinfo {author} {\bibfnamefont
	  {A.}~\bibnamefont {Pasquarello}}, \bibinfo {author} {\bibfnamefont
	  {L.}~\bibnamefont {Paulatto}}, \bibinfo {author} {\bibfnamefont
	  {C.}~\bibnamefont {Sbraccia}}, \bibinfo {author} {\bibfnamefont
	  {S.}~\bibnamefont {Scandolo}}, \bibinfo {author} {\bibfnamefont
	  {G.}~\bibnamefont {Sclauzero}}, \bibinfo {author} {\bibfnamefont {A.~P.}\
	  \bibnamefont {Seitsonen}}, \bibinfo {author} {\bibfnamefont {A.}~\bibnamefont
	  {Smogunov}}, \bibinfo {author} {\bibfnamefont {P.}~\bibnamefont {Umari}},\
	  and\ \bibinfo {author} {\bibfnamefont {R.~M.}\ \bibnamefont {Wentzcovitch}},\
	  }\href {http://www.quantum-espresso.org} {\bibfield  {journal} {\bibinfo
	  {journal} {J. Phys. Condens. Matter}\ }\textbf {\bibinfo {volume} {21}},\
	  \bibinfo {pages} {395502} (\bibinfo {year} {2009})}\BibitemShut {NoStop}%
	\bibitem [{\citenamefont {Kresse}\ and\ \citenamefont
	  {Furthm{\"{u}}ller}(1996)}]{Kresse1996}%
	  \BibitemOpen
	  \bibfield  {author} {\bibinfo {author} {\bibfnamefont {G.}~\bibnamefont
	  {Kresse}}\ and\ \bibinfo {author} {\bibfnamefont {J.}~\bibnamefont
	  {Furthm{\"{u}}ller}},\ }\href {https://doi.org/10.1103/PhysRevB.54.11169}
	  {\bibfield  {journal} {\bibinfo  {journal} {Phys. Rev. B}\ }\textbf {\bibinfo
	  {volume} {54}},\ \bibinfo {pages} {11169} (\bibinfo {year}
	  {1996})}\BibitemShut {NoStop}%
	\bibitem [{\citenamefont {Ponc{\'{e}}}\ \emph {et~al.}(2016)\citenamefont
	  {Ponc{\'{e}}}, \citenamefont {Margine}, \citenamefont {Verdi},\ and\
	  \citenamefont {Giustino}}]{Ponce2016}%
	  \BibitemOpen
	  \bibfield  {author} {\bibinfo {author} {\bibfnamefont {S.}~\bibnamefont
	  {Ponc{\'{e}}}}, \bibinfo {author} {\bibfnamefont {E.~R.}\ \bibnamefont
	  {Margine}}, \bibinfo {author} {\bibfnamefont {C.}~\bibnamefont {Verdi}},\
	  and\ \bibinfo {author} {\bibfnamefont {F.}~\bibnamefont {Giustino}},\ }\href
	  {https://doi.org/10.1016/j.cpc.2016.07.028} {\bibfield  {journal} {\bibinfo
	  {journal} {Comput. Phys. Commun.}\ }\textbf {\bibinfo {volume} {209}},\
	  \bibinfo {pages} {116} (\bibinfo {year} {2016})} \BibitemShut {NoStop}%
	\bibitem {METHOD}%
	The calculations of crystal structures, electronic bandstructures, the projected density of states, Fermi surfaces, and phonons were perform by  Quantum Espresso package. 
	The calculations of Electron-phonon coupling  and related quantities were performed using Quantum Espresso incorporating EPW packages.
	The unfolding bandstructures, unfolding Fermi surfaces, and spin-spiral calculations were performed using VASP package.
	Projector-augmented wave pseudopotentials were chosen, combined with Perdew-Burke-Ernzerhof~\cite{Perdew1996b} incorporating non-local van der Waals interaction (vdw-df2-b86r)~\cite{Lee2010a,Hamada2014}~as exchange-correlation functions. 
	A 15-\AA-thickness vacuum spacing was adopted. 
	The Kohn-Sham valence states were expanded in the plane-wave basis set with a kinetic energy truncation at 50 Ry. 
	The equilibrium crystal structures were determined by a conjugated-gradient relaxation of ionic positions until the Hellmann Feynman force on each atom was  $\textless$0.005 eV/\AA. 
	The samplings of electronic and phonon momenta in BZs were done with two-dimensional $\boldsymbol{k}$- and  $\boldsymbol{q}$-mesh, associated with grid spacing of $2\pi \times 0.017$  and $2\pi \times 0.034$~\AA$^{-1}$, corresponding to grids of 18$\times$18 and 9$\times$9, respectively, whereupon  EPC matrix elements $g_{m n, \nu}(\mathbf{k}, \mathbf{q})$ were calculated~\cite{Ponce2016}, which quantify the scattering amplitude between electronic states ($\boldsymbol{k}$,$m$) and ($\boldsymbol{k}+\boldsymbol{q}$,$n$) via phonon modes ($\boldsymbol{q}$, $\nu$), with $m$($n$) and $\nu$ being electronic and phonon band indexes, respectively. 
	Subsequently, the above quantities were interpolated~\cite{mostofi2008wannier90} onto much denser grids with grid spacing of  $2\pi \times 0.0025$ ($\boldsymbol{k}$-grid) and $2\pi \times 0.005$~\AA$^{-1}$ ($\boldsymbol{q}$-grid). 
	Based on the above grids, the EPC-related quantities, including generalized static electronic susceptibility (\CHIQ), Eliashberg function [$\alpha^{2}F(\omega)$], and momentum-resolved EPC constants ($\lambda_{\boldsymbol{q}\nu}$), were calculated (see main text). 
	The $\delta$ functions were replaced by Gaussian functions in the numerical integration for $\alpha^2F(\omega)$, with a broadening of 30 and 0.5 meV for electrons and phonons, respectively. 
	Unfolded bandstructures and Fermi surfaces were calculated using the  method described in our previous work~\cite{zheng2017charge}, based on the theory introduced in the reference~\cite{Allen2013}.  
	
	\bibitem [{\citenamefont {Perdew}\ \emph {et~al.}(1996)\citenamefont {Perdew},
	  \citenamefont {Burke},\ and\ \citenamefont {Ernzerhof}}]{Perdew1996b}%
	  \BibitemOpen
	  \bibfield  {author} {\bibinfo {author} {\bibfnamefont {J.~P.}\ \bibnamefont
	  {Perdew}}, \bibinfo {author} {\bibfnamefont {K.}~\bibnamefont {Burke}},\ and\
	  \bibinfo {author} {\bibfnamefont {M.}~\bibnamefont {Ernzerhof}},\ }\href
	  {https://doi.org/10.1103/PhysRevLett.77.3865} {\bibfield  {journal} {\bibinfo
	   {journal} {Phys. Rev. Lett.}\ }\textbf {\bibinfo {volume} {77}},\ \bibinfo
	  {pages} {3865} (\bibinfo {year} {1996})}\BibitemShut {NoStop}%
	\bibitem [{\citenamefont {Lee}\ \emph {et~al.}(2010)\citenamefont {Lee},
	  \citenamefont {Murray}, \citenamefont {Kong}, \citenamefont {Lundqvist},\
	  and\ \citenamefont {Langreth}}]{Lee2010a}%
	  \BibitemOpen
	  \bibfield  {author} {\bibinfo {author} {\bibfnamefont {K.}~\bibnamefont
	  {Lee}}, \bibinfo {author} {\bibfnamefont {{\'{E}}.~D.}\ \bibnamefont
	  {Murray}}, \bibinfo {author} {\bibfnamefont {L.}~\bibnamefont {Kong}},
	  \bibinfo {author} {\bibfnamefont {B.~I.}\ \bibnamefont {Lundqvist}},\ and\
	  \bibinfo {author} {\bibfnamefont {D.~C.}\ \bibnamefont {Langreth}},\
	  }\href@noop {} {\bibfield  {journal} {\bibinfo  {journal} {Phys. Rev. B}\
	  }\textbf {\bibinfo {volume} {82}},\ \bibinfo {pages} {81101} (\bibinfo {year}
	  {2010})}\BibitemShut {NoStop}%
	\bibitem [{\citenamefont {Hamada}(2014)}]{Hamada2014}%
	  \BibitemOpen
	  \bibfield  {author} {\bibinfo {author} {\bibfnamefont {I.}~\bibnamefont
	  {Hamada}},\ }\href@noop {} {\bibfield  {journal} {\bibinfo  {journal} {Phys.
	  Rev. B}\ }\textbf {\bibinfo {volume} {89}},\ \bibinfo {pages} {121103}
	  (\bibinfo {year} {2014})}\BibitemShut {NoStop}%
	\bibitem [{\citenamefont {Mostofi}\ \emph {et~al.}(2008)\citenamefont
	  {Mostofi}, \citenamefont {Yates}, \citenamefont {Lee}, \citenamefont {Souza},
	  \citenamefont {Vanderbilt},\ and\ \citenamefont
	  {Marzari}}]{mostofi2008wannier90}%
	  \BibitemOpen
	  \bibfield  {author} {\bibinfo {author} {\bibfnamefont {A.~A.}\ \bibnamefont
	  {Mostofi}}, \bibinfo {author} {\bibfnamefont {J.~R.}\ \bibnamefont {Yates}},
	  \bibinfo {author} {\bibfnamefont {Y.-S.}\ \bibnamefont {Lee}}, \bibinfo
	  {author} {\bibfnamefont {I.}~\bibnamefont {Souza}}, \bibinfo {author}
	  {\bibfnamefont {D.}~\bibnamefont {Vanderbilt}},\ and\ \bibinfo {author}
	  {\bibfnamefont {N.}~\bibnamefont {Marzari}},\ }\href@noop {} {\bibfield
	  {journal} {\bibinfo  {journal} {Comput. Phys. Commun.}\ }\textbf {\bibinfo
	  {volume} {178}},\ \bibinfo {pages} {685} (\bibinfo {year}
	  {2008})}\BibitemShut {NoStop}%
	\bibitem [{\citenamefont {Zheng}\ \emph {et~al.}(2018)\citenamefont {Zheng},
	  \citenamefont {Zhou}, \citenamefont {Liu},\ and\ \citenamefont
	  {Feng}}]{zheng2017charge}%
	  \BibitemOpen
	  \bibfield  {author} {\bibinfo {author} {\bibfnamefont {F.}~\bibnamefont
	  {Zheng}}, \bibinfo {author} {\bibfnamefont {Z.}~\bibnamefont {Zhou}},
	  \bibinfo {author} {\bibfnamefont {X.}~\bibnamefont {Liu}},\ and\ \bibinfo
	  {author} {\bibfnamefont {J.}~\bibnamefont {Feng}},\ }\href@noop {} {\bibfield
	   {journal} {\bibinfo  {journal} {Phys. Rev. B}\ }\textbf {\bibinfo {volume}
	  {97}},\ \bibinfo {pages} {081101(R)} (\bibinfo {year} {2018})}\BibitemShut
	  {NoStop}%
	\bibitem [{\citenamefont {Allen}\ \emph {et~al.}(2013)\citenamefont {Allen},
	  \citenamefont {Berlijn}, \citenamefont {Casavant},\ and\ \citenamefont
	  {Soler}}]{Allen2013}%
	  \BibitemOpen
	  \bibfield  {author} {\bibinfo {author} {\bibfnamefont {P.~B.}\ \bibnamefont
	  {Allen}}, \bibinfo {author} {\bibfnamefont {T.}~\bibnamefont {Berlijn}},
	  \bibinfo {author} {\bibfnamefont {D.~A.}\ \bibnamefont {Casavant}},\ and\
	  \bibinfo {author} {\bibfnamefont {J.~M.}\ \bibnamefont {Soler}},\ }\href
	  {https://doi.org/10.1103/PhysRevB.87.085322} {\bibfield  {journal} {\bibinfo
	  {journal} {Phys. Rev. B}\ }\textbf {\bibinfo {volume} {87}},\ \bibinfo
	  {pages} {239904} (\bibinfo {year} {2013})} \BibitemShut {NoStop}%
	\bibitem [{\citenamefont {Meetsma}\ \emph {et~al.}(1990)\citenamefont
	  {Meetsma}, \citenamefont {Wiegers}, \citenamefont {Haange},\ and\
	  \citenamefont {{De Boer}}}]{Meetsma1990}%
	  \BibitemOpen
	  \bibfield  {author} {\bibinfo {author} {\bibfnamefont {A.}~\bibnamefont
	  {Meetsma}}, \bibinfo {author} {\bibfnamefont {G.~A.}\ \bibnamefont
	  {Wiegers}}, \bibinfo {author} {\bibfnamefont {R.~J.}\ \bibnamefont
	  {Haange}},\ and\ \bibinfo {author} {\bibfnamefont {J.~L.}\ \bibnamefont {{De
	  Boer}}},\ }\href@noop {} {\bibfield  {journal} {\bibinfo  {journal} {Acta
	  Crystallogr. Sect. C Cryst. Struct. Commun.}\ }\textbf {\bibinfo {volume}
	  {46}},\ \bibinfo {pages} {1598} (\bibinfo {year} {1990})}\BibitemShut
	  {NoStop}%
	\bibitem {SM} See Supplemental Materials for the additional computational results supporting the conclusions draw in the main text.
	\bibitem [{\citenamefont {Wang}\ \emph
	  {et~al.}(2022{\natexlab{b}})\citenamefont {Wang}, \citenamefont {Wang},
	  \citenamefont {Feng},\ and\ \citenamefont {Loh}}]{Wang2022a}%
	  \BibitemOpen
	  \bibfield  {author} {\bibinfo {author} {\bibfnamefont {Z.}~\bibnamefont
	  {Wang}}, \bibinfo {author} {\bibfnamefont {Z.}~\bibnamefont {Wang}}, \bibinfo
	  {author} {\bibfnamefont {Y.~P.}\ \bibnamefont {Feng}},\ and\ \bibinfo
	  {author} {\bibfnamefont {K.~P.}\ \bibnamefont {Loh}},\ }\href
	  {https://doi.org/10.1021/acs.nanolett.2c02723} {\bibfield  {journal}
	  {\bibinfo  {journal} {Nano Lett.}\ }\textbf {\bibinfo {volume} {22}},\
	  \bibinfo {pages} {7615} (\bibinfo {year} {2022}{\natexlab{b}})}\BibitemShut
	  {NoStop}%
	\bibitem [{\citenamefont {Piscanec}\ \emph {et~al.}(2004)\citenamefont
	  {Piscanec}, \citenamefont {Lazzeri}, \citenamefont {Mauri}, \citenamefont
	  {Ferrari},\ and\ \citenamefont {Robertson}}]{Piscanec2004}%
	  \BibitemOpen
	  \bibfield  {author} {\bibinfo {author} {\bibfnamefont {S.}~\bibnamefont
	  {Piscanec}}, \bibinfo {author} {\bibfnamefont {M.}~\bibnamefont {Lazzeri}},
	  \bibinfo {author} {\bibfnamefont {F.}~\bibnamefont {Mauri}}, \bibinfo
	  {author} {\bibfnamefont {A.~C.}\ \bibnamefont {Ferrari}},\ and\ \bibinfo
	  {author} {\bibfnamefont {J.}~\bibnamefont {Robertson}},\ }\href
	  {https://doi.org/10.1103/PhysRevLett.93.185503} {\bibfield  {journal}
	  {\bibinfo  {journal} {Phys. Rev. Lett.}\ }\textbf {\bibinfo {volume} {93}},\
	  \bibinfo {pages} {185503} (\bibinfo {year} {2004})} \BibitemShut
	  {NoStop}%
	\bibitem [{\citenamefont {Giustino}(2017)}]{Giustino2017}%
	  \BibitemOpen
	  \bibfield  {author} {\bibinfo {author} {\bibfnamefont {F.}~\bibnamefont
	  {Giustino}},\ }\href {https://doi.org/10.1103/RevModPhys.89.015003}
	  {\bibfield  {journal} {\bibinfo  {journal} {Rev. Mod. Phys.}\ }\textbf
	  {\bibinfo {volume} {89}},\ \bibinfo {pages} {15003} (\bibinfo {year}
	  {2017})}\BibitemShut {NoStop}%
	\bibitem [{\citenamefont {Flicker}\ and\ \citenamefont {{Van
	  Wezel}}(2016)}]{Flicker2016}%
	  \BibitemOpen
	  \bibfield  {author} {\bibinfo {author} {\bibfnamefont {F.}~\bibnamefont
	  {Flicker}}\ and\ \bibinfo {author} {\bibfnamefont {J.}~\bibnamefont {{Van
	  Wezel}}},\ }\href {https://doi.org/10.1103/PhysRevB.94.235135} {\bibfield
	  {journal} {\bibinfo  {journal} {Phys. Rev. B}\ }\textbf {\bibinfo {volume}
	  {94}},\ \bibinfo {pages} {235135} (\bibinfo {year} {2016})}\BibitemShut {NoStop}%
	\bibitem [{\citenamefont {Zhu}\ \emph {et~al.}(2015)\citenamefont {Zhu},
	  \citenamefont {Cao}, \citenamefont {Zhang}, \citenamefont {Plummer},\ and\
	  \citenamefont {Guo}}]{Zhu2015}%
	  \BibitemOpen
	  \bibfield  {author} {\bibinfo {author} {\bibfnamefont {X.}~\bibnamefont
	  {Zhu}}, \bibinfo {author} {\bibfnamefont {Y.}~\bibnamefont {Cao}}, \bibinfo
	  {author} {\bibfnamefont {J.}~\bibnamefont {Zhang}}, \bibinfo {author}
	  {\bibfnamefont {E.~W.}\ \bibnamefont {Plummer}},\ and\ \bibinfo {author}
	  {\bibfnamefont {J.}~\bibnamefont {Guo}},\ }\href
	  {https://doi.org/10.1073/pnas.1424791112} {\bibfield  {journal} {\bibinfo
	  {journal} {Proc. Natl. Acad. Sci. U. S. A.}\ }\textbf {\bibinfo {volume}
	  {112}},\ \bibinfo {pages} {2367} (\bibinfo {year} {2015})} \BibitemShut {NoStop}%
	\bibitem [{\citenamefont {Kohn}(1959)}]{kohn1959image}%
	  \BibitemOpen
	  \bibfield  {author} {\bibinfo {author} {\bibfnamefont {W.}~\bibnamefont
	  {Kohn}},\ }\href@noop {} {\bibfield  {journal} {\bibinfo  {journal} {Phys.
	  Rev. Lett.}\ }\textbf {\bibinfo {volume} {2}},\ \bibinfo {pages} {393}
	  (\bibinfo {year} {1959})}\BibitemShut {NoStop}%
	\bibitem [{\citenamefont {Zheng}\ and\ \citenamefont
	  {Feng}(2019)}]{Zheng2019a}%
	  \BibitemOpen
	  \bibfield  {author} {\bibinfo {author} {\bibfnamefont {F.}~\bibnamefont
	  {Zheng}}\ and\ \bibinfo {author} {\bibfnamefont {J.}~\bibnamefont {Feng}},\
	  }\href {https://doi.org/10.1103/PhysRevB.99.161119} {\bibfield  {journal}
	  {\bibinfo  {journal} {Phys. Rev. B}\ }\textbf {\bibinfo {volume} {99}},\
	  \bibinfo {pages} {161119(R)} (\bibinfo {year} {2019})}\BibitemShut {NoStop}%
	\bibitem [{\citenamefont {McMillan}(1968)}]{McMillan1968}%
	  \BibitemOpen
	  \bibfield  {author} {\bibinfo {author} {\bibfnamefont {W.~L.}\ \bibnamefont
	  {McMillan}},\ }\href {https://doi.org/10.1103/PhysRev.167.331} {\bibfield
	  {journal} {\bibinfo  {journal} {Phys. Rev.}\ }\textbf {\bibinfo {volume}
	  {167}},\ \bibinfo {pages} {331} (\bibinfo {year} {1968})}\BibitemShut
	  {NoStop}%
	\bibitem [{\citenamefont {Allen}\ and\ \citenamefont
	  {Dynes}(1975)}]{Allen1975}%
	  \BibitemOpen
	  \bibfield  {author} {\bibinfo {author} {\bibfnamefont {P.~B.}\ \bibnamefont
	  {Allen}}\ and\ \bibinfo {author} {\bibfnamefont {R.~C.}\ \bibnamefont
	  {Dynes}},\ }\href@noop {} {\bibfield  {journal} {\bibinfo  {journal} {Phys.
	  Rev. B}\ }\textbf {\bibinfo {volume} {12}},\ \bibinfo {pages} {905} (\bibinfo
	  {year} {1975})}\BibitemShut {NoStop}%
	\bibitem {TC}%
	According to the experimental measurement, the \TC~of \TELS~(2.8 K~\cite{Yang2018}) is slightly lower than that of its monolayer counterpart (3.0--3.4 K~\cite{Yang2018}), whose $\lambda$ was calculated to be 0.79~\cite{Lian2022}. 
	Therefore, the $\lambda$ of  \TELS~is expected to be slightly lower than 0.79.
	\bibitem {SIGMA}%
	N. F. Hinsche and K. S. Thygesen, 2D Materials 5, 15009 (2018)
\end{thebibliography}
%

\end{document}